\documentclass[11pt]{article}
\usepackage{graphicx}
\usepackage{subcaption}
\usepackage{tabularray}
\newcommand{\indep}{\rotatebox[origin=c]{90}{$\models$}}
\usepackage{authblk}
\usepackage{subcaption}
\usepackage{blindtext}
\usepackage{graphicx}
\usepackage{amssymb}
\usepackage{multirow}
\usepackage{amsmath}
\usepackage{bm}
\usepackage{amsthm}
\usepackage{float}
\usepackage{cancel}
\usepackage{cite}
\usepackage{comment}
\usepackage{caption}
\usepackage{booktabs,tabularx}
\usepackage{adjustbox}
\usepackage{pgfplotstable}
\newcommand{\subtitle}[1]{%
	\posttitle{%
		\par\end{center}
	\begin{center}\large#1\end{center}
	\vskip0.5em}%
	}
	\date{}
	\usepackage{float}
	\usepackage{multirow}
	\usepackage{pdfpages}
	\usepackage{fullpage}
	\usepackage{verbatim}
	\usepackage{cite}
	\usepackage[english]{babel}
	\usepackage{float}
	\usepackage[hidelinks=true]{hyperref}
	\hypersetup{
colorlinks   = true, 
urlcolor     = blue, 
linkcolor    = blue, 
citecolor    = blue 
}

\usepackage{color}
\usepackage[hmargin=1in, inner=1in,outer=1in,bottom=1in,top=1in]{geometry}
\usepackage{setspace}
\usepackage[all]{hypcap}
\usepackage[square,sort,comma,numbers]{natbib}
\usepackage{amsmath,amsfonts}
\usepackage{grffile}
\usepackage[all]{xy}
\usepackage{url}

\title{Estimation of the Number Needed to Treat, the Number Needed to be Exposed, and the Exposure Impact Number with Instrumental Variables}
\author{Valentin Vancak$^{1,2}$\footnote{Corresponding author: Valentin Vancak, Department of Data Science, Holon Institute of Technology, Holon, Israel, E-mail: valentin.vancak@gmail.com}\, \& Arvid Sjölander$^2$} 

\affil{$^1$Department of Data Science, Holon Institute of Technology, Holon, Israel\\ 
	\vspace{0.3cm}
	$^2$Department of Medical Epidemiology and Biostatistics,
Karolinska Institutet,
Stockholm, Sweden}

\begin{document}

\maketitle

\abstract{
	The number needed to treat (NNT) is an efficacy index defined as the average number of patients needed to treat to attain one additional treatment benefit. In observational studies, specifically in epidemiology, the adequacy of the populationwise NNT is questionable since the exposed group characteristics may substantially differ from the unexposed. To address this issue, groupwise efficacy indices were defined: the Exposure Impact Number (EIN) for the exposed group and the  Number Needed to be Exposed (NNE) for the unexposed. Each defined index answers a unique research question since it targets a unique sub-population. In observational studies, the group allocation is typically affected by confounders that might be unmeasured. The available estimation methods that rely either on randomization or the sufficiency of the measured covariates for confounding control will result in inconsistent estimators of the true EIN, NNE, and NNT. Using Rubin's potential outcomes framework, we explicitly define the NNT and its derived indices as causal contrasts. Next, we introduce a novel method that uses instrumental variables to estimate the three aforementioned indices in observational studies. We present two analytical examples and a corresponding simulation study. The simulation study illustrates that the novel estimators are consistent, unlike the previously available methods, and their confidence intervals meet the nominal coverage rates. Finally, a real-world data example of the effect of vitamin D deficiency on the mortality rate is presented.\\
	\\
$\mathbf{Keywords}:$ NNT, NNE, EIN, Instrumental variables, G-estimation 
}

\newpage 
\section{Introduction}\label{sec:NNT}
The Number Needed to Treat (NNT) is a widely used efficacy index in the analysis of randomized controlled trials (RCTs), as well as in epidemiology and meta-analyses.\cite{newcombe2012confidence,  vancak2021guide, lee2020cost, verbeek2019cost, da2012methods, mendes2017number}  However, in observational studies, the characteristics of the exposed group can differ substantially from those of the unexposed group. To address this, group-specific efficacy indices have been defined: the Exposure Impact Number (EIN) for the exposed group and the Number Needed to be Exposed (NNE) for the unexposed group. While the NNT serves as a populationwise index, the EIN and NNE provide group-specific insights. Consequently, these three indices support different decisions and address different research questions. 

The NNT provides a straightforward measure of a treatment's effectiveness in preventing a negative outcome (harm) or achieving a positive one (benefit) in a population. Originally, the NNT was defined as the average number of patients who need to be treated to prevent one additional adverse event.\cite{laupacis1988assessment, kristiansen2002number}  Another common definition of the NNT is the number of patients who need to be treated to achieve one additional positive outcome due to the treatment, termed as the treatment benefit. These definitions are fundamentally equivalent, as preventing an adverse event can be viewed as achieving a beneficial outcome. In epidemiological contexts, this concept is adapted to refer to ``exposure benefit,'' reflecting the positive outcome due to exposure rather than treatment.

In epidemiological studies, using and interpreting the marginal populationwise NNT becomes challenging due to the inherent differences between the treated and untreated groups. Unlike RCTs, observational studies often deal with populations where the treated and untreated groups may differ significantly in their baseline characteristics. This discrepancy can lead to misleading interpretations if the populationwise NNT is applied without an appropriate adjustment. To address this, group-specific efficacy indices have been introduced: the EIN for the treated (exposed) group and the NNE for the untreated (unexposed) group.\cite{bender2002calculating} These indices provide more accurate measures in observational studies, reflecting the effectiveness of treatment (exposure) within the treated (exposed) and the untreated (unexposed) subpopulations, respectively. From now on, the term \textit{treatment} is replaced with the term \textit{exposure} to be consistent with the common terminology in observational studies.

Each of these indices - NNT, EIN, and NNE - answers a unique research question. The populationwise NNT is most relevant when considering interventions intended for  an entire population. In contrast, the EIN and NNE are particularly useful in scenarios where the exposure is optional. These distinctive measures allow for a more nuanced understanding of exposure effects in observational studies, facilitating more accurate and meaningful interpretations of exposure efficacy within different subpopulations. In observational studies, the group allocation is typically affected by confounders. The available estimation methods of the EIN, NNE, and NNT assume that the measured confounders are sufficient for confounding control or that the group allocation was randomized. Such assumptions are generally not reasonable in observational studies. Therefore, such estimation methods will produce statistically inconsistent estimators that may lead to distorted or even completely inadequate conclusions and subsequent decisions. 

The main contribution of this study is providing an explicit causal formulation of the EIN, NNE,
and NNT alongside a comprehensive theoretical framework for their point and interval estimation in observational studies with possible unmeasured confounders. This advancement is critical for improving the estimation accuracy of these indices in real-world settings. Our approach enhances the reliability and usefulness of these measures, offering statistically consistent estimates that are essential for informed decision-making in public health and clinical practice.

\section{Background and notations}

Several authors \citep{bender2007estimating, bender2010estimating, sjolander2018estimation, mueller2022personalized, vancak2022number} have noticed that the causal meaning of the NNT and its derived indices (EIN, NNE) is embedded in their very definition. Hence, we will use Rubin's potential outcomes framework\cite{rubin2005causal} to define all three indices. Since any of these indices is a one-to-one mapping of the exposure benefit in the corresponding (sub)population, we start with its formal definition. Let~$Y_1$ be the potential outcome for a given individual if exposed, and~$Y_0$ be the potential outcome for the same individual if the individual was not exposed. Let~$A$ be the exposure indicator where its realization is denoted by the subscript~$a$, such that $a=1$ denotes exposure and $a=0$ non-exposure. Let~$I_a$ be the potential dichotomous outcome if the exposure~$A$ is set to $a$, $a\in \{0, 1\}$. If the potential  outcome is binary, then $I_a \equiv Y_a$, $a \in \{0, 1\}$. Otherwise, a dichotomization is performed, e.g., $I_a \equiv I\{Y_a \le \tau\}$, for $a\in \{0, 1\}$. The formal definition of the exposure benefit is $\{I_1 = 1, I_0 = 0\}$. Namely, a benefit that is caused by the exposure such that without the exposure, no benefit will occur. Therefore, a possible formal definition of the exposure benefit~$p_b$ is\cite{mueller2022personalized} 
\begin{align}\label{eq:p_b}
	P(I_1 = 1, I_0 = 0).
\end{align}
Another common definition of the exposure benefit~$p_b$ is   
\begin{align}\label{eq:ATE}
	E[I_1 - I_0] = P(I_1 = 1) - P(I_0=1)	.
\end{align}	
This quantity is also known as the average treatment effect (ATE). However, it is qualitatively different from~$p_b$ as defined in eq.~\eqref{eq:p_b}. Particularly, one can expand $p_1 = P(I_1 = 1) = P(I_1 = 1, I_0 = 0) + P(I_1 = 1, I_0 = 1)$, and $p_0 = P(I_0 = 1) = P(I_1 = 1, I_0 = 1) + P(I_1 = 0, I_0 = 1)$. Therefore, the ATE can be expressed as 
\begin{align*}
	p_1 - p_0 =  p_b - P(I_1 = 0, I_0=1).  
\end{align*}
Namely, whenever the treatment can be harmful, i.e., $P(I_1 = 0, I_0=1) > 0$. In such a case, the support set of ATE is $[-1,1]$. Therefore, it cannot be interpreted as a probability. Several authors~\cite{mueller2022personalized, pearl2022probabilities, schulzer1996unqualified} defined the NNT (EIN, NNE) explicitly as the reciprocal number of the exposure benefit probability~$P(I_1 = 1, I_0 = 0)$. Other authors~\cite{laubender2010estimating, walter2001number}  refrained from explicitly defining the NNT as the inverse of a probability; however, they assumed that $p_1 \ge p_0$. This assumption restricts the ATE to $[0,1]$, and thus it can be interpreted as a probability, even if it was not explicitly done so. However, it is still not the same probability as in eq.~\eqref{eq:p_b} since it allows for the exposure to be harmful for a certain subpopulation, as long as it is beneficial on average. In order to unify the different approaches, one can use the monotonicity assumption.\cite{pearl2022probabilities, angrist1996identification, angrist1995two} Formally, assuming $I_1 \ge I_0$, for any individual in the target population. In other words, assuming that the treatment is not harmful on the individual level.\footnote{For further discussion and numerical examples, please refer to Mueller and Pearl.\cite{mueller2022personalized}} In such a case, the ATE is non-negative and equals the exposure benefit probability~$P(I_1 = 1, I_0 = 0)$. However, the monotonicity assumption induces unnecessary restriction without introducing any substantial advantages. Without the monotonicity assumption, the target parameter remains the ATE. However, in such a case, the interpretation of the estimates is different. To wit, relaxing the monotonicity assumption changes the interpretation of the target parameter but does not affect the presented methodology. Therefore, we do not assume monotonicity and thus estimate the appropriate ATE, which is the exposure benefit as defined in eq.~\eqref{eq:ATE}, for each of the indices.

In order to intuitively understand the NNT, let~$N$ be the number of exposed individuals. Thus,~$NE [I_1] = Np_1$ is the number of beneficial outcomes if all~$N$ individuals are exposed. Analogically, $NE[I_0] = Np_0$ is the number of beneficial outcomes if all~$N$ individuals are not exposed. Therefore, the NNT is defined as the quantity that solves the following equation for~$N$
\begin{align*}
	N(p_1 - p_0) = 1.
\end{align*} 
Alternatively, assume a random variable~$N_b$ that counts the number of exposed individuals out of an infinitely large super-population until the first exposure benefit occurs. Then~$N_b$ follows geometric distribution in which the expected value is known to be
\begin{align*}
	E[N_b] = \frac{1}{p_b} = \frac{1}{p_1 - p_0}.
\end{align*} 
Namely, $E[N_b]$ is the expected number of people to be exposed to observe the first exposure benefit. Using these two motivational examples, we arrive at the same original Laupcais' et al.\cite{laupacis1988assessment} definition of the NNT. Assuming that the exposure is beneficial on average implies $p_1 - p_0 \equiv p_b > 0$. Due to sampling variability, this definition may result in negative point estimators of the NNT. Such estimators lead to difficulties in their interpretation,\cite{grieve2003number, hutton2000number, snapinn2011clinical, sonbo2004cost, kristiansen2002number} bi-modal sample distribution,\cite{grieve2003number} and infinite disjoint confidence intervals~(CIs). All these non-desirable properties occur due to singularity at~$0$ of the original definition. Vancak et al.\cite{vancak2020systematic} resolved the pitfall of singularity at~$0$ by modifying the original definition of the NNT. The modified NNT is
\begin{align}\label{def.g}
	\text{NNT} \equiv g(p_b) = 
	\begin{cases}
		1/p_b, \quad p_b >0 \\
		\infty ,\quad \,\,\,\, \,  p_b \le 0 \, . 
	\end{cases}
\end{align}   
We adopt this modification for all three indices; EIN, NNE, and NNT. Namely, these indices are defined by applying the function~$g$~ as in eq.\eqref{def.g} to the corresponding exposure benefit.

The fundamental problem of causal inference is that it is impossible to observe both~$I_1$ and~$I_0$ within the same individual,\cite{holland1986statistics} as the individual is either exposed or unexposed. Formally, $I = I_0I\{A=0\} + I_1I\{A=1\}$. The counterfactuals-based definition highlights the main challenge with the three indices - their estimation. The parameter~$p_b$ cannot be estimated directly since we cannot distinguish between a successful outcome due to exposure (i.e., exposure benefit) and a successful outcome not due to exposure (i.e., non-exposure benefit) on an individual level. Therefore, a straightforward practice to estimate $p_b$ is using randomized controlled trials (RCTs) where the exposure group is used to estimate~$p_1$, and the control (unexposed) group to estimate~$p_0$.  

\noindent In order to define the EIN and NNE, we need first define the general exposure benefit as a function of the exposure indicator~$A$ 
\begin{align}\label{eq:general_p}
	p_b(A)  = E[I_1 - I_0 | A],
\end{align}
which is the conditional ATE. The marginal exposure benefit~$p_b$ is obtained by $E[p_b(A)]$.  To define the groupwise exposure benefit~$p_b(a)$, we set the exposure~$A$ in eq.~\eqref{eq:general_p} to a fixed value~$a$. Namely,
$
p_b(a) = E[I_1 - I_0 | A=a],
$
$a\in \{0, 1\}$, where~$p_b(0)$ and~$p_b(1)$ are the exposure benefits for the unexposed and the exposed groups, respectively. Consequently, the EIN is defined as $g(p_b(1))$ and the NNE as $g(p_b(0))$.  If the exposure is randomized, then $p_b(0) = p_b(1) = p_b$, and  EIN = NNE = NNT. Otherwise, generally, $p_b(0) \neq p_b(1)$, and the marginal exposure benefit is  their weighted mean, i.e., $p_b = E[p_b(A)]$. 
Next, we define the general conditional exposure benefit. Let~$L$ be background characteristics, such as age, sex, and other relevant features measured in the study baseline. Therefore, the general conditional exposure benefit is
\begin{align}\label{eq:general_cond_p}
	p_b(A, L) =  E[I_1 - I_0 | A, L].
\end{align}
The conditional exposure benefit for the $a$th group $p_b(a;L)$ is obtained by setting the exposure indicator~$A$ to a fixed value~$a$ in eq.~\eqref{eq:general_cond_p}, i.e., $p_b(a;L) = p_b(A=a,L)$.  
The groupwise marginal exposure benefits~$p_b(0)$, $p_b(1)$, can be obtained by averaging the groupwise conditional exposure benefit over~$L$ in the corresponding group. Consequently, the true EIN and NNE are obtained by $g(E[ p_b(1;L)|A=1])$ and $g(E[ p_b(0;L)|A=0])$, respectively. Analogically, the true NNT is obtained by $g(E[ p_b(A, L)])$. 

\begin{figure}
	\[
	\xymatrixrowsep{0.4cm}
	\xymatrixcolsep{1cm}
	\xymatrix{ &&& U \ar[dl] \ar[dr] &\\
		Z \ar[rr]&&  A \ar[rr] && I  \\
		&& L \ar[llu] \ar[u] \ar[rru] &&
	}
	\]
	\caption{\small{A causal structure of a valid IV.~$I$ is the outcome variable, $A$ is exposure, and~$Z$ is the instrument. $U$ represents all unmeasured confounders of~$A$ and~$I$, whereas~$L$ represents all measured confounders. The instrument~$Z$ affects~$I$ only through the exposure~$A$, and is not confounded with the outcome by the unmeasured variables.}   }\label{dag:valid_iv}
\end{figure}
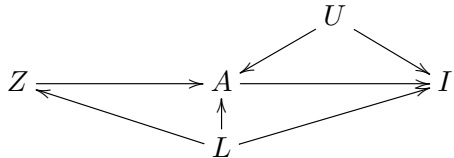

The available estimation methods rely either on randomization of exposure allocation or on sufficiency of~$L$ for confounding control.\cite{vancak2022number, sjolander2018estimation, bender2002calculating} Formally, in non-randomized settings, the counterfactual part~$E[I_a|A = 1 - a]$ of the groupwise exposure benefit~$p_b(a)$ is identified as 
\begin{align*}
	E[I_a|A = 1 -  a] = E[E[I| A = a, L] | A = 1 - a],
\end{align*}
only if~$L$ is assumed to be sufficient for confounding control. This assumption is not reasonable  in most observational studies. 
In this article, we focus on estimating the marginal EIN, NNE, and NNT in observational studies where there are no measured confounders or the measured confounders~$L$ are insufficient for confounding control. The proposed estimation method relies on instrumental variables (IVs), which, under appropriate assumptions, allows for consistent estimation of the three indices. An IV~$Z$ is a variable that is associated with the exposure~$A$, affects the outcome~$I$ only through the exposure, and is not confounded with the outcome by unmeasured confounders~$U$. Please refer to Figure~\ref{dag:valid_iv} for a graphical illustration of an IV. The use of IV regression methods is a common practice in epidemiological studies. However, to the authors' knowledge, IV regression has not yet been used to estimate the EIN, NNE, or NNT. 
A possible application of the two-stage least-squares (TSLS)\cite{angrist1995two} method is restrictive since it requires linear models with continuous outcomes and exposure. Namely, the TSLS regression requires specific assumptions on the original latent data-generating process prior to dichotomization. However, since the outcome is assumed to be dichotomous for the NNT (NNE, EIN) calculations, the TSLS is not always feasible in such settings since the estimation must be performed on the original data scale, which is not always available. On the other hand, our method is based on modeling the already dichotomous outcome's causal structure, whether the outcome was dichotomized or is naturally dichotomous. Such an approach aligns with the usual workflow where the data analysis starts after possible dichotomization. Notably, this practice has disadvantages - dichotomization results in a loss of Fisher information and, therefore, a decrease in statistical power.\cite{senn2003disappointing, fedorov2009consequences} However, dichotomization is unavoidable if the researcher uses the NNT (NNE, EIN) indices on non-dichotomous outcome. Whether the dichotomization is justified is usually a clinical question, which is out of the scope of our study. Unlike the TSLS method, our approach allows the structural model to consider arbitrary link functions and does not require assumptions on possible latent processes. This makes the novel method more appealing and applicable to various applications.

The rest of the paper is structured as follows: Section~\ref{sec2} defines and models the exposure benefit in observational studies using generalized structural mean models. Section~\ref{sec3} provides two illustrative examples of the model presented earlier. In particular, we introduce the double logit and the double probit models. Section~\ref{sec4} introduces the novel estimation method. In Section~\ref{sec5}, we conduct a simulation study to compare the new estimation method to the previously available methods. Section~\ref{sec6} presents an analysis of real-world data. In this analysis, we use the aforementioned indices to measure the effect of vitamin D deficiency on mortality. Section~\ref{sec7} concludes the study by summarizing its main results and discussing future research directions.

\section{Modeling the exposure benefit}\label{sec2}

\subsection{Instrumental variables} 
Let~$\{L, U\}$ be a set of all confounders of the exposure~$A$ and the outcome~$I$, such that~$L$ denotes measured confounders and~$U$ denotes unmeasured confounders. In scenarios where $U =\emptyset$, $L$ is sufficient for confounding control.   However, in many observational studies $U\neq \emptyset$, thus the measured confounders~$L$ are insufficient for confounding control. In such a case, the estimation of the conditional and marginal exposure benefits by adjusting for the measured confounders~$L$ will result in inconsistent estimators of the exposure benefits and their corresponding efficacy indices. A possible approach to address the issue of confounding bias to use instrumental variables (IVs). Formally, an IV~$Z$ is a variable that is (1) relevant, i.e., is associated with exposure~$A$, (2) exogenous, i.e., is not associated with unmeasured confounders, and (3) affects the outcome~$Y$ only through the exposure~$A$ (exclusion). For graphical illustration of a valid IV please refer to Figure~\ref{dag:valid_iv}. The exclusion and the exogeneity assumptions can be compactly formulated using counterfactuals: $I_a \indep Z | L$, $a \in \{1, 0\}$.\cite{didelez2010assumptions} For a graphical illustration of the counterfactual formulation of a valid IV, please refer to Figure~\ref{dag:twin_net}. In the following subsection, we introduce generalized structural mean models that are used to model the exposure benefits and the corresponding indices. Since we assume that~$L$ is insufficient for confounding control, the observed and the potential outcomes are conditionally associated with the instrument~$Z$, given exposure~$A$. This occurs since conditioning on the exposure opens a non-causal path from the (potential) outcome to~$Z$ through the unmeasured confounder~$U$, on which~$A$ is a collider. Therefore, both the causal and the associative models below are conditioned on~$Z$. 

\subsection{Generalized structural mean model}
Robins\cite{robins1989analysis, robins1994correcting} introduced a new class of additive and multiplicative structural mean models to estimate the ATE on the treated for continuous and positive outcomes in scenarios with unmeasured confounders. However, these models suffered from the inherent limitations of the linear models used for modelling binary outcomes. For example, the inability to secure predicted outcomes that are either dichotomous or lie between~$0$ and~$1$. Therefore, Vansteelandt and  Goetghebeur\cite{vansteelandt2003causal} generalized the former structural models. The generalized structural mean models (GSMM) are semiparametric two-stage models that can handle, among other settings, binary outcomes. The first stage model is the generalized structural model indexed with causal parameters that convey the causal effects of interest. The second stage model is the association generalized model for the observed data that is required to identify the causal parameters of the former. The structural and association models' two link functions can differ. However, in such a case, possible issues are introduced when there are no unmeasured confounders. In such a  scenario, the parameters of the association model equal the causal parameters. Therefore, the two models need to coincide.  Notably, as the association model models the observed data, all its features can be empirically tested. At the same time, the structural model is a latent model defined using counterfactuals. Thus, its features cannot be empirically tested without additional restricting assumptions. Throughout this study, we assume that the working models are correctly specified. In particular, we construct the presented examples so that model compatibility is guaranteed through the chosen values of the association model parameters.

The GSMM of the first stage specifies the mean causal effect for the exposed (unexposed) individuals. Formally, let $\xi$ be a strictly monotone and continuously differentiable link function.  We assume that the GSMM is  
\begin{align}\label{eq:gsmm}
	\xi\left(E[I_1 | Z, L, A=a]\right)   
	-
	\xi\left(E[I_0 | Z, L, A=a]\right)
	=
	m_a^T(L) \psi_a , \quad a \in \{0, 1\}, 
\end{align} 
where $\psi_a \in \Psi_a$ is the vector of causal parameters for the $a$th group, $a \in \{0, 1\}$, and $\text{dim}(m_a(L)) = \text{dim}(\psi_a)$. The composition of the vector-valued function~$m_a(L)$ defines the exact form of the causal model and may allow for interactions between~$L$ and the exposure~$A$ in the $a$th group. In addition, since the causal parameters also depend on the group~$a$, it allows for a distinct model for each exposure group~$a \in \{0, 1\}$. According to the consistency assumption, $E[I_a | Z, L, A=a] = E[I | Z, L, A=a]$.\cite{didelez2010assumptions} Therefore, this part of the model is identifiable from the observed data. However, the identification of the counterfactual part $E[I_a | L, Z, A = 1 - a ]$ requires a model and a valid IV. Particularly,  the conditional independence of~$I_a$ and~$Z$ given~$L$ implies the  mean independence as presented in the following equality   
\begin{align}\label{valid_iv}
	E[I_a | Z, L ] = E[I_a | L ],\quad a \in \{0, 1\}.
\end{align}  
To estimate the conditional benefit, we need to specify the second stage model - a statistical association model for the observed outcomes. Let~$\eta$ be another link function, then the association model is defined as $\eta\left(E[I | Z,  L, A  ]\right)$. Hence, the counterfactual quantity $E[I_a | L, Z, A = 1 - a ]$ can be expressed as 
\begin{align}\label{eq:counter_predict}
	E[I_a | L, Z, A = 1 - a ] = \xi^{-1} \left\{
	\eta \left(  E[I|Z, L, A= a  ]\right)  - a m_1 ^ T(L) \psi_1 + (1-a) m^T_0\psi_0 
	\right\}, \quad a \in \{0, 1\}.
\end{align}
If $\eta$ is a link function of a generalized linear model, then the association model can be expressed as a linear function indexed with parameters~$\beta$. Therefore, the counterfactual part of the structural model can be expressed as a function of the observed data and the unknown parameters~$\psi$ and~$\beta$. 
If $L$ is insufficient for confounding control, conditioning on the exposure $A$ opens a non-causal path between the observed and the potential outcomes and the instrument~$Z$ (please refer to Figure~\ref{dag:twin_net} for illustration). This leads to a notable distinction between the association and the structural model. The structural mean model, which models the causal effect of the exposure on the exposed (unexposed) individuals, is a contrast between the mean of two potential outcomes. Therefore, although unmeasured confounders $U$, given the exposure $A=a$, induce a non-causal dependency between the potential outcome $I_a$ and the instrument $Z$, this dependency cancels out in the contrast.  However, in the association model, we model directly the mean dependency of the observed outcome $I$ on the exposure, the measured confounders and the instrument. Therefore, the association model's dependence on the instrument $Z$ must be explicitly modelled.   
Generally, the two link functions might be different. For example, if~$\eta$ is the probit link function, and~$\xi$ is the logit link function, we have a probit-logit joint model. However, in many scenarios, one may assume that $\xi = \eta$.  Two-stage GSMMs with the same link function~$\xi$ for the structural and the association model are referred to as double~$\xi$ models, where~$\xi$ is replaced with the relevant link function. 

\subsection{Modeling of the exposure benefit}
The form of the general conditional exposure benefit~\eqref{eq:general_cond_p} is
\begin{align}\label{eq:cond_exp_pb_A}
	p_b(Z, L, A) = 	E[I_1 - I_0 | Z, L, A].
\end{align}
Since this quantity depends on the exposure~$A$, to obtain the conditional populationwise exposure benefit, we average it over the conditional distribution of~$A$ given~$Z$ and~$L$,
\begin{align}\label{eq:cond_pb}
	p_b(Z, L) = 	E[p_b(Z, L, A)|Z, L].
\end{align} 
Consequently, the marginal populationwise exposure benefit~$p_b$ is obtained by averaging $p_b(Z, L)$ over the joint distribution of $Z$ and $L$, i.e., 
\begin{align}\label{eq:marginal_pb}
	p_b = 	E[p_b(Z, L)] = E[I_1 - I_0].
\end{align}
Thus, the populationwise NNT can be obtained by applying~$g$ tp~$p_b$, i.e., $g(p_b)$.  The conditional exposure benefit for the~$a$th group is obtained by setting the exposure indicator~$A$ to a fixed value~$a$ in eq.~\eqref{eq:cond_exp_pb_A}, i.e., 
\begin{align}\label{eq:cond_p_b(a)}
	p_b(a;Z, L) = 	E[I_1 - I_0| Z, L, A=a], \quad a \in \{0, 1\}. 
\end{align}	
By calculating the expectation w.r.t. the joint conditional distribution of~$Z$, $L$, in the~$a$th group, one can recover the marginal exposure benefit in the~$a$th group, and consequently the EIN and the NNE. Namely, the groupwise exposure benefit is given by
\begin{align}\label{eq:marginal_pb(a)}
	p_b(a) = E[p_b(a;Z, L )], \quad a \in\{0, 1\},
\end{align}
where the EIN and NNE are obtained by applying~$g$ to $p_b(1)$ and $p_b(0)$, respectively. 
Assuming that $\xi = \eta$, the general conditional exposure benefit can be formulated as 
\begin{align}\label{eq:exp_prob}
	p_b(Z, L, A, \psi ) 
	&= 
	\xi ^ {-1} 
	\left(
	\xi( E[I|Z, L, A] )	
	+  m^T_{0}(L)\psi_{0}(1-A) )
	\right)\\
	& -
	\xi ^ {-1} 
	\left(
	\xi( E[I|Z, L, A ])	
	- m^T_{1}(L)\psi_{1} A 
	\right). 	\nonumber 
\end{align}
The conditional exposure benefit for the $a$th group, $p_b(a;Z, L, \psi_a )$, is obtained by setting the exposure~$A$ to a fixed value~$a$ in eq.~\eqref{eq:exp_prob}, namely, 	$p_b(a;Z, L, \psi_a ) 
= 
p_b(Z, L, a, \psi)$,  $a\in \{0, 1\}$.
The exact form of the conditional populationwise and groupwise exposure benefits  depend on the GSMM and the specified association model. 

\section{Example}\label{sec3}
\subsection{The double logit model}  
Assume a binary outcome $I \in \{0, 1\}$, a binary exposure $A \in \{0, 1\}$, and a binary instrument~$Z \in \{0, 1\}$. Assume there are no measured confounders, i.e., $L = \emptyset$ and $m(L) = 1$. 
For the logit link function~$\xi$, we define the GSMM~\eqref{eq:gsmm} as    
\begin{align}\label{eq:logit_causal_model}
	\text{logit} \left( E[ I_1  = 1 | Z, A = a ]  \right)  - \text{logit} \left( E[ I_0  = 1 | Z, A = a] \right)  = \psi_a, \quad a \in \{0, 1\}, 
\end{align}
where $\text{logit}(x) = \ln\left(\frac{x}{1-x}\right)$. Notably, there are two causal parameters $\psi = (\psi_0, \psi_1)^T \in \Psi$, where $\Psi \equiv \Psi_0 \cup \Psi_1$, which represent the mean causal effect of the exposure on the unexposed and the exposed, respectively. Additionally, assume the following saturated association logit model for the observed outcome
\begin{align}\label{eq:logit_assoc_mod}
	\text{logit} ( E[I|Z, A;\beta_I ] ) = \beta_0 + \beta_1A + \beta_2 Z + \beta_3 ZA.
\end{align}
The vector of coefficients $\beta_I  = (\beta_0, \beta_1, \beta_2, \beta_3)^T \in \mathcal{B}$ incorporates the association between~$Z, A$ and~$I$ which, at least partially, results from the omission of~$U$. Particularly, if $U = \emptyset$, then $\beta_2 = \beta_3 = 0$ since $I\indep Z|L$; thus, $\beta_1 = \psi_0 = \psi_1$. Otherwise, if $U \neq \emptyset$, then $\beta_1 \neq \psi_a$, and~$\psi_a$, $a \in \{0, 1\}$, cannot be estimated using the classical maximum likelihood method. However, the association model coefficients~$\beta_I$ can still be estimated using the score equations that result from the maximum likelihood approach. Using the GSMM and the logit association model, we can express the general conditional exposure benefit~\eqref{eq:exp_prob} as a function of~$\beta_I$ and~$\psi$
\begin{align}\label{eq:logit_exp_prob}
	p_b(Z,  A, \beta_I, \psi ) 
	&= 
	\text{expit} 
	\left(
	\beta_0 + \beta_1A + \beta_2Z + \beta_3ZA	
	+  \psi_{0}(1-A) )
	\right)\\
	& -
	\text{expit} 
	\left(
	\beta_0 + \beta_1A + \beta_2Z + \beta_3ZA		
	- \psi_{1} A 
	\right), 	\nonumber 
\end{align}
where $\text{expit}(x) = ( 1 - \text{exp}(-x) ) ^ {-1}$. To compute the populationwise NNT, we apply the function~$g$ to the marginal populationwise exposure benefit~$p_b = E[p_b(Z, A,  \beta_I,\psi)]$~\eqref{eq:marginal_pb}, i.e.,  $\text{NNT} = g(p_b)$, where the expectation is taken w.r.t. the joint distribution of~$A$ and~$Z$. If we set~$A$ to~$0$ in eq.~\eqref{eq:logit_exp_prob} we obtain the conditional exposure benefit~\eqref{eq:cond_p_b(a)} for the unexposed  $a=0$ as a function of $\beta_I$ and $\psi_0$ 
\begin{align}\label{eq:logit_p_unexposed}
	p_b(0; Z, \beta_I, \psi_0) =  \text{expit} ( \beta_0 + \beta_2Z + \psi_0)
	- 
	\text{expit}( \beta_0 + \beta_2Z).
\end{align} 
To compute the NNE, we apply the function~$g$ to the marginal exposure benefit in the unexposed group $p_b(0) = E[p_b(0; Z, \beta_I, \psi_0)|A=0]$~\eqref{eq:marginal_pb(a)} \eqref{eq:logit_p_unexposed}, i.e., $\text{NNE} = g(p_b(0))$.   Analogically, for the conditional exposure benefit~\eqref{eq:cond_p_b(a)} for the exposed $a=1$, we set~$A$ to~$1$ in eq.~\eqref{eq:logit_exp_prob}
\begin{align}\label{eq:logit_p_exposed}
	p_b(1; Z, \beta_I, \psi_1)  & = \text{expit}(\beta_0 + \beta_1 + \beta_2 Z + \beta_3 Z)
	- \text{expit}(\beta_0 + \beta_1 + \beta_2Z + \beta_3 Z - \psi_1).
\end{align}
To compute the EIN, we apply the function~$g$ to the marginal exposure benefit  in the exposed group~$p_b(1) = E[p_b(1; Z, \beta_I, \psi_1)|A=1]$~\eqref{eq:marginal_pb(a)} \eqref{eq:logit_p_unexposed}, i.e., $\text{EIN} = g(p_b(1))$.
Notably, for a binary exposure~$A$, and binary instrument~$Z$, any probabilistic model that models the distribution of~$A$ given~$Z$ is saturated; therefore, no additional parametrizations are required.  By repeating similar steps for the probit~$\Phi^{-1}$ link function~$\xi$, we can obtain the explicit forms of the conditional and the marginal exposure benefits with the corresponding indices for the double probit model. Please refer to Appendix~\ref{app:double_probit_example} for the explicit derivations.

\subsection{Remark}
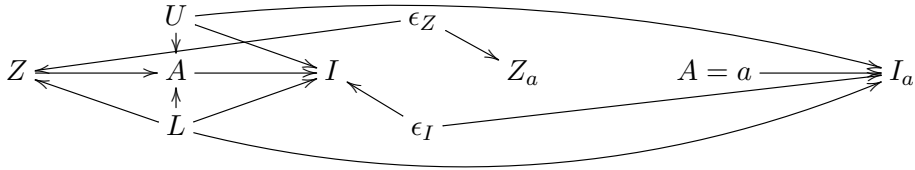
\begin{figure}
	\[
	\xymatrixrowsep{0.2cm}
	\xymatrixcolsep{0.7cm}
	\xymatrix{ && U  \ar[d] \ar[drr]  \ar@/^1.1pc/[rrrrrrrrd] &&& \ar[dlllll] \epsilon_Z \ar[dr] &&&&&  &\\
		Z \ar[rr]&& A \ar[rr] && I &&  Z_a&& A = a \ar[rr]&& I_a\\
		&& L \ar[llu] \ar[u] \ar[rru] \ar@/_2pc/[rrrrrrrru] && &  \epsilon_I \ar[lu]  \ar[rrrrru]\\
		&&&&&
	}
	\]
	\caption{\small{DAG of a twin network causal model with instrumental variable. The left-hand side of the DAG represents the observed actual world, while the right-hand side represents the hypothetical potential world. On the left-hand side, $Z$ is the IV, $A$ is the exposure, $L$ is the set of measured confounders, $U$ are the unmeasured confounders, and~$I$ is the outcome. On the right-hand side, $Z_a$ is the potential value of the IV where $A$ is set to $a$. Since $Z$ is not affected by the exposure, $Z_a = Z$. In addition, $A=a$ is the specified value of the exposure, and $I_a$ is the potential outcome under this exposure.  For both DAGs, $\epsilon_Z$ and $\epsilon_I$ represent all the unmeasured exogenous factors that determine the values of~$Z$ and $I$, respectively.}  }\label{dag:twin_net}
\end{figure}

If there is no causal effect of the exposure on the outcome, i.e., $\psi_a = 0$, then $p_b(a; Z, \beta_I, 0) = 0$, $a\in \{0, 1\}$; namely, the exposure benefit is~$0$, and EIN=NNE=NNT=$\infty$. If there is non-zero causal effect $\psi_a$, $a\in \{0, 1\}$, and the exposure is randomized, then $\beta_1 = \psi_0 = \psi_1$, and $\beta_2 = \beta_3 = 0$ since there is no open path between~$Z$ and~$I_a$ (please refer to the DAG of a twin causal network in Figure~\ref{dag:twin_net} for illustration). In such a case, the three efficacy indices coincide and are equal to~$g(p_b)$. Moreover, in such a scenario, the average exposure benefit reduces to the observable difference
\begin{align*}
	E[I|A=1] - E[I|A=0],
\end{align*}
that can be readily estimated by the difference between the corresponding sample means. If we have measured confounders~$L$, the  steps above can be repeated by conditioning on~$L$. Finally, the exposure benefits in both examples are functions of unknown parameters~$\beta_I$ and~$\psi$. The association model coefficients~$\beta_I$ can be estimated using, for example, score functions. However, for the causal parameters~$\psi$, a different approach is required. The next section introduced the G-estimation method for estimating the causal parameter~$\psi$ and the corresponding indices.

\section{The G-estimator}\label{sec4}

The SMMs were introduced with a consistent estimation methodology called the G-estimation. Robins developed the G-estimation for the linear and log link functions.\cite{robins1989analysis,robins1994correcting} Vansteelandt and Goetghebeur\cite{vansteelandt2003causal} later extended the theory by developing a G-estimator for the GSMM. The underlying idea of the G-estimation approach is to combine the structural and association models to construct a prediction of the counterfactual outcome for each subject. Then, this prediction and an auxiliary function of the instrument~$Z$ are used to construct estimating equations.\cite{stefanski2002calculus} The values of $\psi = (\psi_0, \psi_1)^T$ that solve these estimating equations are called the G-estimators. Loosely speaking, the G-estimators are the values of the causal parameters~$\psi$ under which a valid IV assumption $I_a \indep Z|L$, $a \in \{0, 1\}$ holds in the observed data. Notably, the G-estimators coincide with the traditional TSLS in the special scenario where the structural and association models are linear with continuous outcome and exposure. However, since the outcome is necessarily dichotomous in scenarios where the EIN, NNE, and NNT  are computed, the traditional TSLS estimators can be viewed only as approximations that generally yield asymptotically biased estimators.\cite{vansteelandt2011instrumental} 
Vansteelandt and  Goetghebeur\cite{vansteelandt2003causal} derived asymptotically normal and efficient G-estimators of the structural model causal parameters. In the scenario with no exposure effect, they derived an estimator that is robust  to misspecification of the association model. However, in other scenarios, a correct specification of the association model is required since  its misspecification  can invalidate the null hypothesis tests  on the ATE.  To overcome this limitation, Robins and Rotnitzky\cite{robins2004estimation} proposed an alternative method that avoids direct specification of the association model; however, it requires  the specification of three additional parametric models for the conditional distribution of the exposure~$A$, the mean potential outcome~$I_a$, and an additional modified structural mean model, such that these models are always compatible with the first stage structural model. Therefore, to estimate the causal parameters~$\psi$ and subsequently the corresponding EIN, NNE, and NNT, we use the G-estimators for the GSMM introduced by Vansteelandt and  Goetghebeur.\cite{vansteelandt2003causal}

Assume that the association model of the observed outcome follows a parametric structure indexed by the vector of coefficients~$\beta_I$, i.e., $\xi (E[I|Z, L, A;\beta_I])$. For construction of the estimating equations, we define a function~$h(a; Z, L, A, \beta_I, \psi_a)$ for estimation of the counterfactual mean $E[I_{1-a}|Z, L, A=a]$. Particularly,  for a double~$\xi$ model, we define a function that is generalized version of the function in eq.~\eqref{eq:counter_predict}
\begin{align}\label{def:h_a}
	h(a; Z, L, A, \beta_I, \psi ) = 
	\xi^{-1}\{ \xi ( E[I|Z, L, A; \beta_I] ) - 
	am_{1}^T(L) \psi_{1}A   +    
	(1-a) m_{0}^T(L) \psi_{0}(1-A)\}, \,\, a\in \{0, 1\}. 
\end{align}
Eq.~\eqref{def:h_a} defines two distinct functions: one for the unexposed group $a=0$, and another for the exposed group $a=1$. The explicit form of the function in eq.~\eqref{def:h_a} for the double logit and the double probit models are obtained by replacing~$\xi$ with the logit and probit functions, respectively. For further construction of the estimating equations, we define an auxiliary function $D_a(Z, L)$, $a\in \{0, 1\}$, which is an arbitrary function with the same dimension as~$\psi_a$, that satisfies $E[D_a( Z, L)|L] = 0$. For a one-dimensional instrument~$Z$, we define $D_a(Z, L) = m_a(L)(Z - E[Z|L])$, where~$E[Z|L]$ is the instrument model that need to be specified. For example, assume that the instrument model follows a parametric structure with parameters vector~$\pi_Z$, i.e., $E[Z|L ; \pi_Z]$. For more details on the construction of the function $D_a(Z, L)$ and its alternative forms, please refer to Sjölander \& Martinussen.\cite{sjolander2019instrumental}
Finally, the G-estimators of the vector of causal parameters~$\psi = (\psi_0, \psi_1) ^T$ are values that solve the corresponding estimating equations, namely, 
\begin{align}\label{eq:dh}
	\sum_{i=1}^{n} 
	\begin{pmatrix}
		D_{0}(Z_{i}, L_{i}, \pi_Z)   h(0;     Z_i, L_i, A_i, \beta_I, \psi_0)\\
		D_{1}(Z_{i}, L_{i}, \pi_Z) h(1;   Z_i, L_i, A_i, \beta_I, \psi_{1})
	\end{pmatrix}   
	= 0.
\end{align} 
The proof of the consistency of these G-estimators appears in the Appendix~\ref{app:cons_g}.  Notably, these estimating equations depend on the unknown parameters~$\beta_I$, and~$\pi_Z$, which also need to be estimated. Additionally, since our target parameters are NNE, EIN, and NNT, which are defined by applying the function~$g$ to $p_b(0)$, $p_b(1)$, and $p_b$, respectively, we need to extend further eq.~\eqref{eq:dh} to a larger system of estimating equations that incorporates all the parameters above.  Let $\theta$ be the vector of all estimands $(\beta_I, \psi, \pi_Z, p_b(0), p_b(1), p_b, \text{NNE}, \text{EIN}, \text{NNT})^T$. Therefore, the estimating vector-valued function is 
\begin{align}\label{eq:est_fun}
	\mathbf{Q} (L, Z, A; \theta ) =
	\begin{pmatrix}
		\mathbf{S}(Z, L, A; \beta_I, \pi_Z)\\
		\mathbf{Dh}(Z, L, A;  \beta_I,\pi_Z, \psi)\\
		\mathbf{p}\left(Z, L, A; \beta_I, \psi, p_b(0), p_b(1), p_b \right)\\
		\mathbf{g}\left(p_b(0), p_b(1), p_b, \text{NNE, EIN, NNT}\right)\\
	\end{pmatrix}.
\end{align}
and the corresponding estimating equations are
\begin{align}\label{eq:est_eq}
	\sum_{i=1}^n \mathbf{Q} ( Z_i, L_i, A_i; \theta ) = 0.
\end{align}
The vector-valued function  $\mathbf{S}(Z, L, A; \beta_I, \pi_Z )$ is  a vector of unbiased estimating functions of~$\beta_I$ and~$\pi_Z$, 
respectively. Namely,
\begin{align}\label{eq:est_fun_b}
	(S(Z, L, A; \beta_I) , S(Z, L ; \pi_Z))^T. 
\end{align}
The vector-valued function  $\mathbf{Dh}(Z, L, A; \beta_I, \pi_Z, \psi)$ are the estimating functions for the causal parameters~$\psi$ as presented in eq.~\eqref{eq:dh}, i.e., 
\begin{align*}
	\begin{pmatrix}
		D_{0}(Z, L, \pi_Z)   h(0;     Z, L, A, \beta_I, \psi_0), 
		D_{1}(Z, L, \pi_Z)   h(1;   Z, L, A, \beta_I, \psi_{1})
	\end{pmatrix}  ^ T.
\end{align*} 
The vector-valued function $ \mathbf{p}\left(Z, L, A; \beta_I, \psi, p_b(0), p_b(1), p_b \right)$ is 
\begin{align}\label{eq:est_fun_p}
	((p_b(0; Z, L,  \beta_I,  \psi_0) - p_b(0))(1-A),
	(p_b(1; Z, L, \beta_I, \psi_1) - p_b(1))A,
	p_b(Z, L, A, \beta_I, \psi) - p_b  ) ^T,	
\end{align}
where $p_b(0; Z, L, \beta_I,  \psi_0)$ and $p_b(1; Z, L, \beta_I, \psi_1)$ are the conditional groupwise exposure benefits as defined in eq.~\eqref{eq:cond_p_b(a)} for $a=0$ and $a=1$, respectively, and $p_b(Z, L, A, \beta_I, \psi)$ is the conditional general exposure benefit as defined in eq.~\eqref{eq:exp_prob}.  Finally, the vector-valued function  $ \mathbf{g}\left(p_b(0), p_b(1), p_b, \text{NNE, EIN, NNT}\right)$ is  
\begin{align}\label{eq:est_fun_g}
	(g(p_b(0)) - \text{NNE},
	g(p_b(1)) - \text{EIN},
	g(p_b ) - \text{NNT}  ) ^T,
\end{align}
which is required for estimating the corresponding indices. Notably, although this function is independent of the observed data, it is still required for computing the asymptotic variance of the estimators of the NNE, EIN, and NNT, respectively. The asymptotic variance of~$\theta$'s estimators~$\hat \theta$ is obtained by applying the sandwich formula
\begin{align}\label{eq:sadwitch}
	\mathrm{Var}(\hat \theta) = n^{-1} \mathbf{A}(\theta)^{-1} \mathbf{B}(\theta) \mathbf{A} (\theta), 
\end{align} 
where the “bread” matrix is $\mathbf{A}(\theta) = E[-\partial \mathbf Q(\theta)/\partial \theta^{T}]$, the “meat” matrix is $\mathbf{B}(\theta) = E[\mathbf Q(\theta) \mathbf Q(\theta)^T]$, and $\mathbf{Q}(\theta)$ is a shorthand for the vector valued estimating function from eq.~\eqref{eq:est_fun}. The asymptotic distribution of the estimators~$\hat{\theta}$ is multivariate normal
\begin{align*}
	\sqrt{n} (\hat \theta - \theta ) \xrightarrow{D} \mathcal{N}_p(0, \mathrm{Var}(\theta)), 
\end{align*}  
where the subscript~$p$ denotes the dimension of the parametric space~$\theta \in \Theta$, and the superscript~$D$ denotes convergence in distribution.  For the sample version of the “bread” $\mathbf A(\theta)$ and the “meat” $\mathbf B(\theta)$ matrices, we replace the expectation operator with the corresponding sample means, and $\theta$ with ts estimator~$\hat \theta$.

\section{Simulation study}\label{sec5}
To construct the data generating process (DGP) of the observed outcome~$I$ without explicitly using the unmeasured confounders~$U$ in the simulation set-up, we start by specifying the GSMM using causal parameters~$\psi$, the parametric models for the distribution of a valid IV~$Z$, and the exposure~$A$. Since omission of~$U$ affects the values of associative parameters~$\beta_I$ of the outcome model $E[I|Z, L, A ;\beta_I]$, we need to account for this effect when we set their values. In other words, the unmeasured confounders are implicit, whereas the consequence of their omission is incorporated in the values of~$\beta_I$. Moreover,~$\beta_I$ also depend on the values of the causal parameters~$\psi$. Particularly, the regression coefficient~$\beta_1$ corresponding to the exposure indicator~$A$ does not equal the main causal effect~$\psi_1$ of the exposure on the outcome. In order to specify the values of~$\beta_I$ for the DGP, we use the assumed GSMM and explicit forms of valid IV's imposed restrictions~\eqref{valid_iv} on the parametric space. Additionally, we incorporate restrictions imposed by the specification of the outcome's~$I$ marginal distribution and the specification of the marginal exposure benefit~$p_b$. These conditions restrict the parametric space~$\Theta$ and result in a system of non-linear equations, which is solved w.r.t.~$\beta_I$. Since these restrictions propagate to the exposure benefits,   the set of possible NNTs (EINs, NNEs) is also restricted.   The next subsection presents in detail the four-step simulation procedure. The simulation code and the generated data sets are available online on the author's Github repository.\footnote{Simulations source code and generated data sets: \url{https://github.com/vancak/nne_iv}.  }     

\subsection{Simulation setup}\label{subsec:sim_setup}
\begin{figure}
	\begin{subfigure}{0.5\textwidth}
		\centering
		\includegraphics[width=1\linewidth]{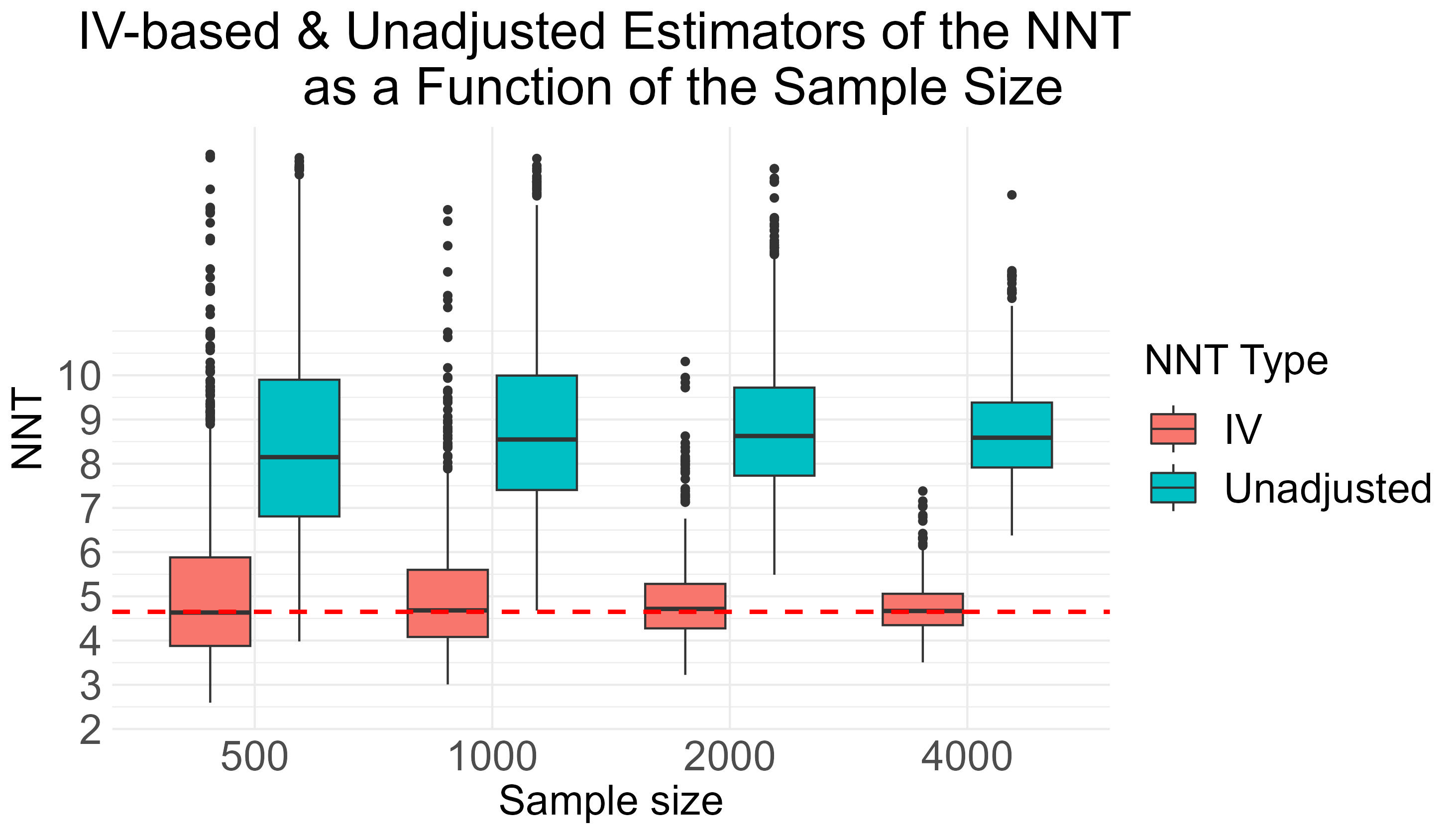}
		\caption{Double logit model. True NNT$= 4.65$.}
	\end{subfigure}%
	\begin{subfigure}{0.51\textwidth}
		\centering
		\includegraphics[width=1\linewidth]{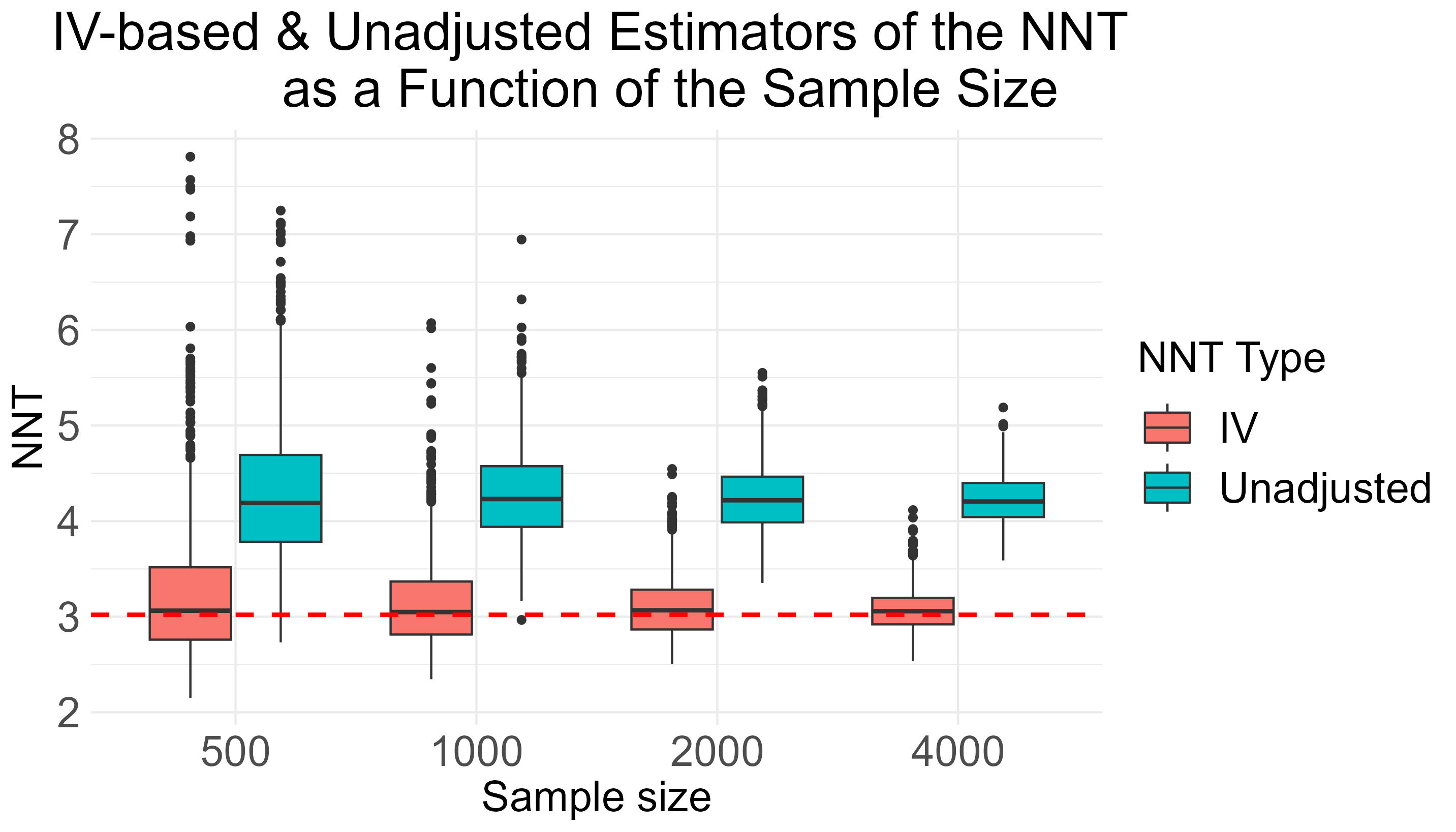}
		\caption{Double probit model. True NNT$= 3.02$.}
	\end{subfigure}
	\caption{Logit and probit models examples: Binary outcome with logit and probit causal models, respectively. The true causal parameters for both models are $\psi = (1, 1.5)^T$. The true populationwise NNTs are $4.65$, and $3.02$, respectively.  The red boxplots denote the IV-based estimators of the NNT, which are based on G-estimators of~$\psi$. The blue boxplots denote the unadjusted estimators of the populationwise NNT. The red dashed line denotes the true NNT value for each model. The calculations were repeated~$m=1000$ times for four different sample sizes: $n = 500, 1000, 2000, 4000$. For both models, the marginal $P(A=1)=0.6$, $\pi_Z = 0.5$, $\gamma = (-0.83, 3)^T$ and the marginal probability of the outcome is $P(I=1) = 0.3$.}\label{sim:nnt_boxplots}
\end{figure}

\begin{figure}
	\begin{subfigure}{0.5\textwidth}
		\centering
		\includegraphics[width=1\linewidth]{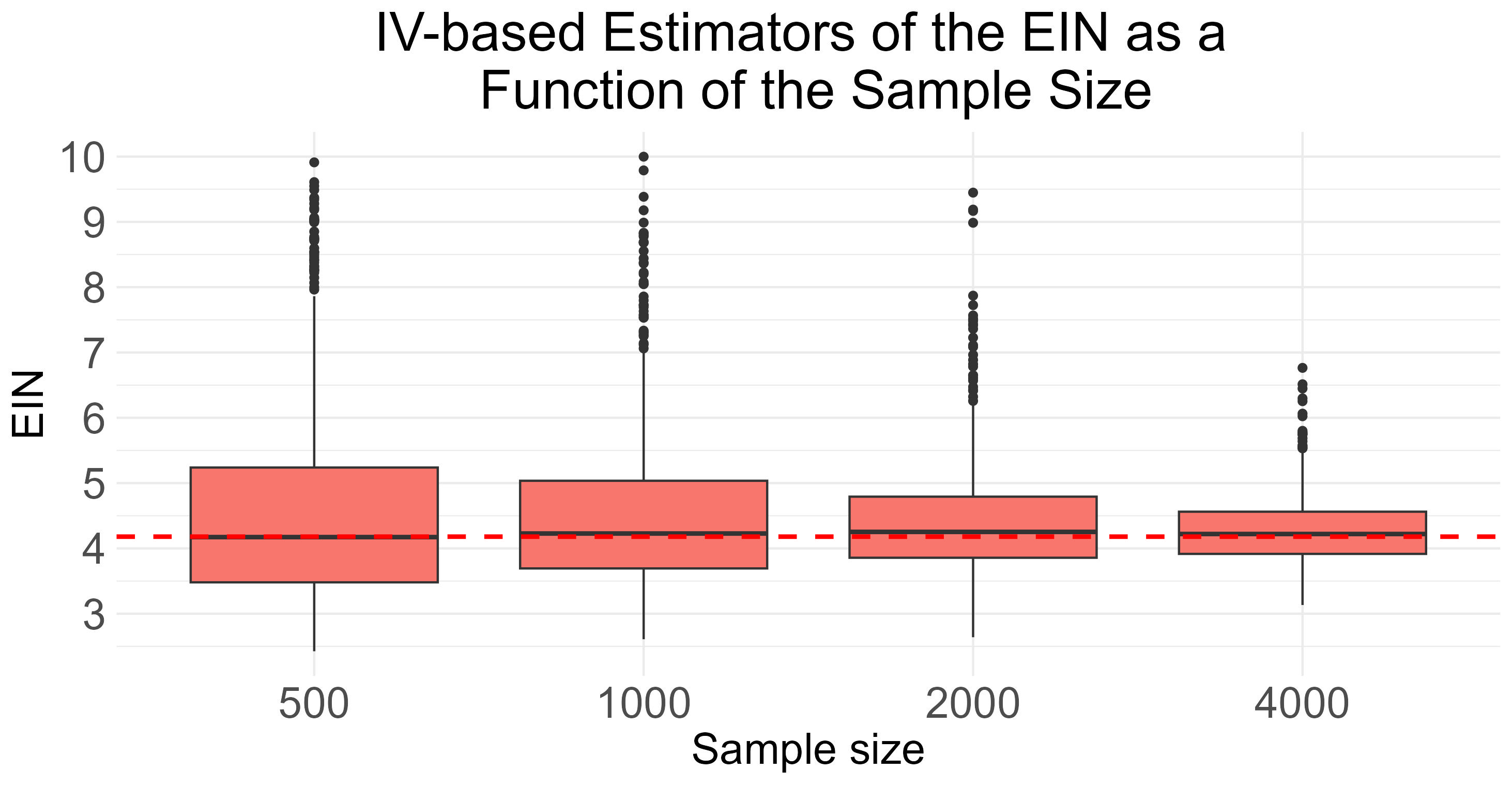}
		\caption{Double logit model. True EIN$= 4.18$.}
	\end{subfigure}%
	\begin{subfigure}{0.5\textwidth}
		\centering
		\includegraphics[width=1\linewidth]{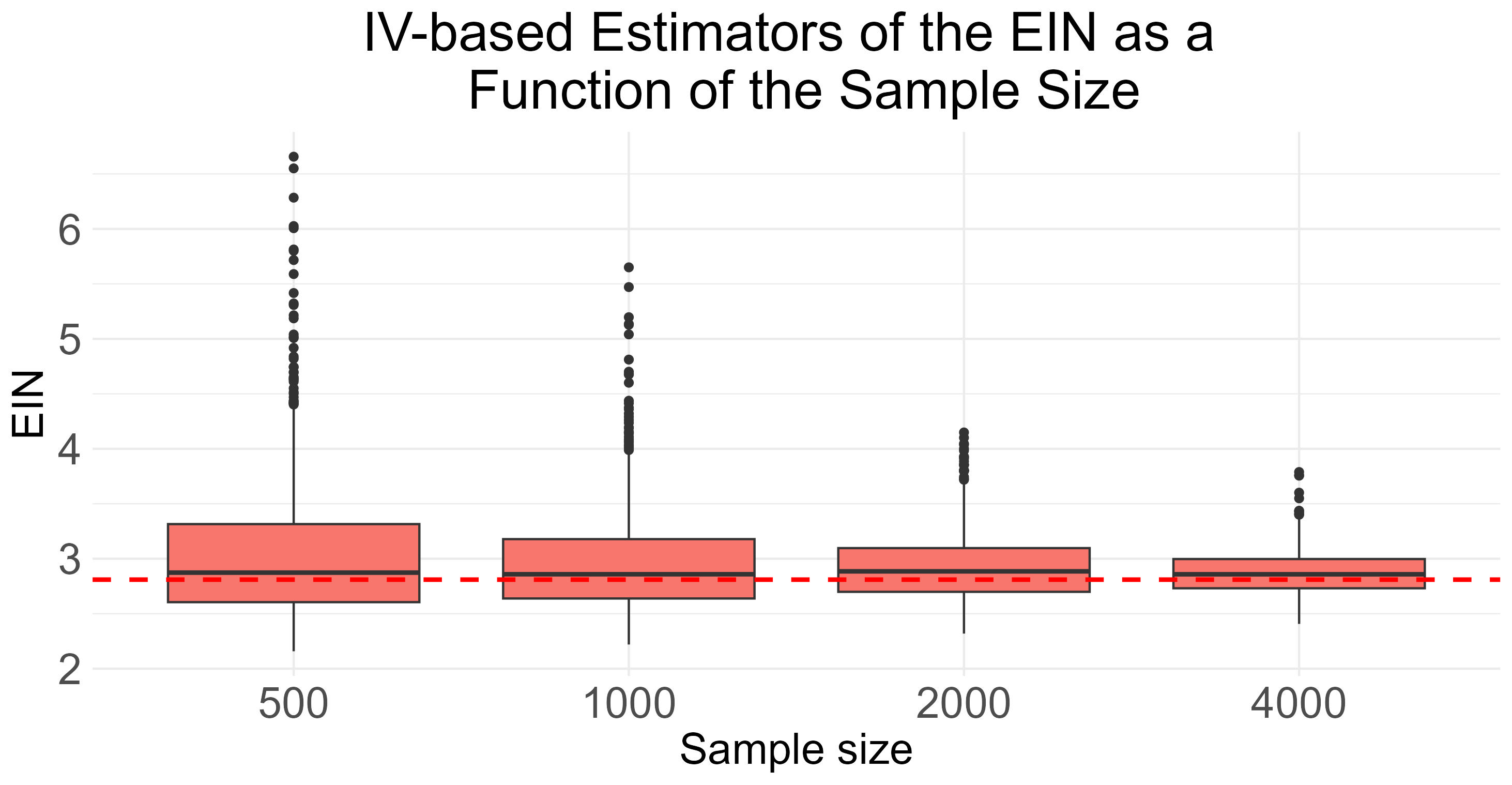}
		\caption{Double probit model. True EIN$= 2.81$.}
	\end{subfigure}
	\caption{\small{Double logit and probit models examples: Binary outcome with logit and probit causal models, respectively. The causal parameters are $\psi = (1, 1.5)^T$. The true EINs are $4.18$ and $2.81$, respectively.  The red boxplots denote the IV-based estimators of the EIN, which are based on G-estimators of~$\psi$. The unadjusted estimators of the EIN were infinitely large (the estimated benefits were negative before the application of the function~$g$~\eqref{def.g}); therefore, they were omitted from the graph.  The red dashed line denotes the true EIN value in each model. The calculations were repeated~$m=1000$ times for four different sample sizes: $n = 500, 1000, 2000, 4000$. For both models, the marginal $P(A=1)=0.6$, $\pi_Z = 0.5$, $\gamma = (-0.83, 3)^T$ and the marginal probability of the outcome is $P(I=1) = 0.3$.}  }\label{sim:ein_boxplots}
\end{figure}

\begin{figure}
	\begin{subfigure}{0.5\textwidth}
		\centering
		\includegraphics[width=1\linewidth]{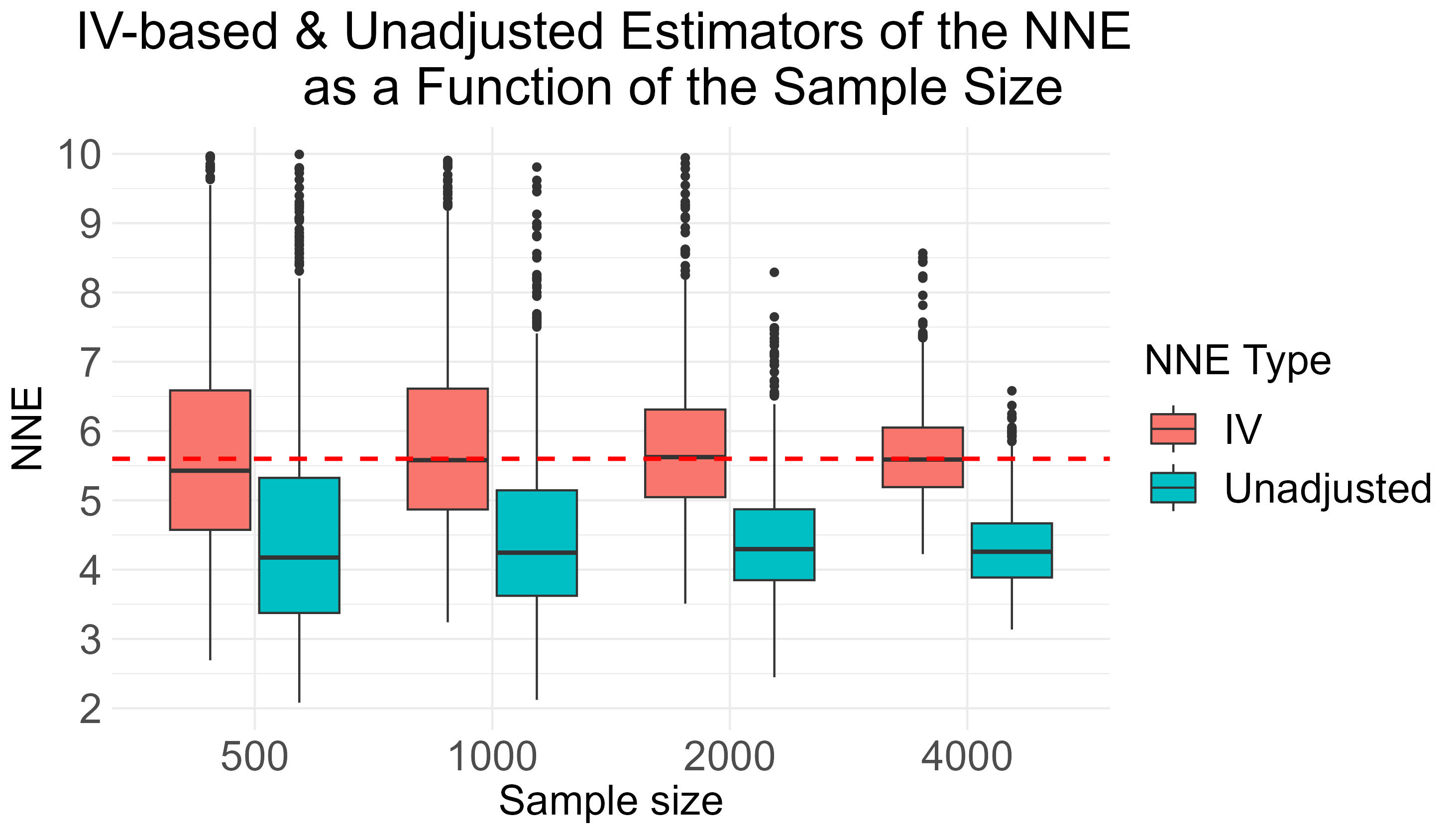}
		\caption{Double logit model. True NNE$= 5.60$.}
	\end{subfigure}%
	\begin{subfigure}{0.5\textwidth}
		\centering
		\includegraphics[width=1\linewidth]{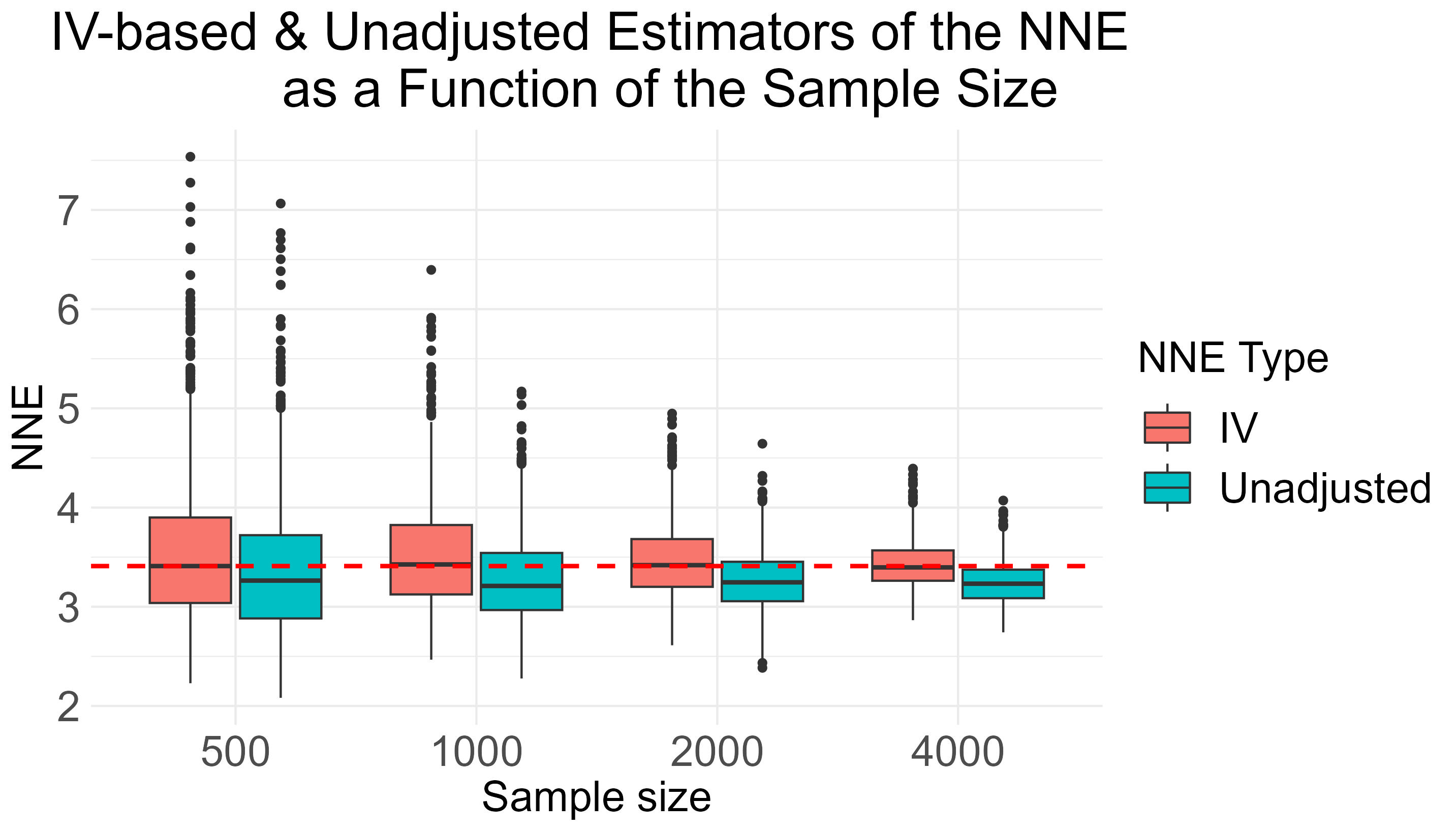}
		\caption{Double probit model. True NNE$= 3.41$.}
	\end{subfigure}
	\caption{
		\small{Double logit and probit models examples: Binary outcome with logit and probit causal models, respectively. The true causal parameters for both models are $\psi = (1, 1.5)^T$. The true populationwise NNEs are $5.60$, and $3.41$, respectively.  The red boxplots denote the IV-based estimators of the NNE, which are based on G-estimators of~$\psi$. The blue boxplots denote the unadjusted estimators of the NNE. The red dashed line denotes the true NNE value for each model. The calculations were repeated~$m=1000$ times for four different sample sizes: $n =  500, 1000, 2000, 4000$. For both models, the marginal $P(A=1)=0.6$, $\pi_Z = 0.5$, $\gamma = (-0.83, 3)^T$ and the marginal probability of the outcome is $P(I=1) = 0.3$.}  }\label{sim:nne_boxplots}
\end{figure}

In the first step, we set the values of the simulation parameters. In the second step, we generate the data. In the third step, we use the data to estimate the target parameters. In the fourth step we assess the efficiency of our method. For both models we assume a binary outcome $I \in \{0, 1\}$, a binary exposure $A \in \{0, 1\}$, and a binary instrument~$Z \in \{0, 1\}$. Additionally, we assume  that there are no measured confounders, i.e., $L = \emptyset$ and $m_a(L) = 1$, $a\in \{0, 1\}$. 
\begin{enumerate}
	\item We set the marginal exposure to~$0.6$, i.e., $P(A=1) = 0.6$, and the marginal outcome probability~$P(I=1)$ to~$0.3$. Next, we specify the conditional distribution of the exposure given the instrument, $P(A|Z; \gamma)$, by specifying the parameters~$\gamma$.
	Formally, 
	\begin{align}\label{dgp:Z&A}
		Z & \sim Ber(\pi_Z) \\
		A | Z & \sim Ber(\text{expit}(\gamma_0 + \gamma_1 Z)).\nonumber
	\end{align}
	We set $\pi_Z = 0.5$, and $\gamma = (-0.83, 3)^T$ to satisfy the aforementioned marginal distributions of the outcome~$I$, and the exposure~$A$. Next, we specify the association model of the observed outcome  
	\begin{align}\label{dgp:outcome_model}
		I|A, Z & \sim Ber(\xi^{-1}(\beta_0 + \beta_1 A + \beta_2 Z + \beta_3 AZ )) ,	
	\end{align}
	where $\xi^{-1}$ is the expit and the inverse probit $\Phi^{-1}$ functions for the double logit and double probit models, respectively. To  set the coefficients values of the association model~\eqref{dgp:outcome_model}, we use the GSMM as described in eq.~\eqref{eq:logit_causal_model}, and eq. \eqref{eq:probit_causal_model} for the logit and the probit link functions, respectively. Particularly, we set the values of the causal parameters to~$\psi = (1, 1.5)^T$ for both models. Next, we set the value of the marginal exposure benefit~$p_b$ to a possible value that satisfies the restrictions above. To specify~$\beta_I$, we use the explicit forms of valid IV conditions $E[I_a|Z] = E[I_a]$, $a\in \{0, 1\}$. For a binary instrument~$Z$, the valid IV conditions boil down to
	\begin{align}\label{eq:valid_iv_binaryZ}
		P( I_a  = 1 | Z = 1) =  P( I_a  = 1 | Z = 0), \quad    a\in \{0, 1\}.
	\end{align}
	The explicit forms of eq.~\eqref{eq:valid_iv_binaryZ}, for $a \in \{0, 1\}$, are given in the Appendix~\ref{app:valid_iv}. Eventually, we solve a system of non-linear equations w.r.t.~$\beta_I$.  With the obtained solution, we compute the groupwise conditional exposure benefit using eq.~\eqref{eq:cond_p_b(a)} for $a\in \{0,1\}$, and the general exposure benefit using eq.~\eqref{eq:exp_prob}. At the end of the first step, we obtain the true values of all parameters of interest~$\theta$. 
	
	\item We use the marginal distribution of the IV~$Z$, and the conditional exposure distribution~$P(A|Z; \gamma)$ as described in~\eqref{dgp:Z&A} to generate $n = 500, 1000, 2000, 4000$ realizations of the instrument~$Z$ and the exposure~$A$, respectively. Next, we use the association model~\eqref{dgp:outcome_model} to generate realizations of the outcome~$I$. These steps are repeated for $m=1000$ times for each sample size~$n$. At the end of each iteration, we have a data set for further analysis. 
	
	\item  For each iteration of Step~2, we estimate all parameters of interest~$\theta$ and their covariance matrix. The point estimators are obtained by solving the system of estimating equations presented in eq.~\eqref{eq:est_eq}. In order to compute the $95\%$-level CIs, we use the sandwich formula as in eq.~\eqref{eq:sadwitch} to compute the empirical covariance matrix. The ``bread'' matrix, which is the minus of the Jacobian matrix of $\mathbf{Q} (L, Z, A; \theta )$~\eqref{eq:est_fun} is computed numerically using the $\mathrm{pracma}$ package in R.\cite{borchers2022package} The ``meat'' matrix computed numerically as well, using the base package in $\mathrm{R}$. The sandwich matrix's main diagonal entries are the corresponding estimands' variances. Therefore, we used the last three entries of the main diagonal corresponding to the computed variances of NNE, EIN, and NNT, respectively. For more details on the estimations procedure, please refer to the Appendix~\ref{app:sim_step2}. At the end of the third step, we have a matrix of~$m$ estimators with corresponding CIs for each sample size~$n$.

	\item  In the final fourth step, we compare the IV-based estimators with the unadjusted (for unmeasured confounders) estimators in terms of statistical consistency (see Figures~\ref{sim:nnt_boxplots}, \ref{sim:ein_boxplots}, and \ref{sim:nne_boxplots}). Additionally, we compute the empirical coverage rates of the $95\%$-level CIs, the Monte Carlo standard errors (MCSEs) and the average bias of the estimators (see Table~\ref{tb:coverage_rate}).   
	
\end{enumerate}

\subsection{Simulation summary}
Graphical summary of the estimators' behaviour as a function of the sample size~$n$ can be found in Figures~\ref{sim:nnt_boxplots}, ~\ref{sim:ein_boxplots}, and~\ref{sim:nne_boxplots}. Figure~\ref{sim:nnt_boxplots} illustrate the behaviour of the populationwise NNTs estimators, Figure~\ref{sim:ein_boxplots} illustrate the behaviour of the EINs estimators, and Figure~\ref{sim:nne_boxplots} illustrate the behaviour of the NNEs estimators for the double logit and probit models, respectively. Notably, since~$g$, as defined in eq.~\eqref{def.g}, is a strictly convex transformation of the exposure benefit, the empirical distributions of all the estimators exhibit right-skewness. This skewness results from applying the convex transformation~$g$ to the approximately symmetric distribution of the estimators  of the exposure benefit~$p_b$. Additionally, one can observe that for the populationwise NNT, the unadjusted estimators are biased upwards, whereas for the EIN and NNE, they are biased downwards. For the EIN, the bias is severe and produces infinitely large estimators, while for the NNE, the bias is rather small. For all indices, the estimators of the logit model are less stable (have larger standard errors compared to the probit model). 

Table~\ref{tb:coverage_rate} presents the empirical coverage rates of the $95\%$-level CI, the MCSEs and the average bias for the marginal EIN, NNE, and NNT in double logit and probit models as a function of the sample size~$n$. The empirical coverage rates are close to the nominal $95\%$ for the double logit model and the NNE index for the double probit model. The CIs overshoot the nominal coverage rate for the double probit model for EIN and NNT indices.  This behaviour  was also reflected in the proportion of non-informatively extremely large CIs for EIN and NNT for sample sizes of $500$ and $1000$ (13.5\%, 8\%, and 10.6\%, 5.8\% for EIN and NNT, respectively).
Additionally, one can see that the smaller the sample size, the weaker the IV. Therefore, the estimators for such sample sizes are less stable, and their average bias is higher.\cite{burgess2017review} For all examined settings, the average bias is larger than the corresponding MCSEs, decreasing drastically as the sample size increases. This behaviour illustrates the statistical consistency of the IV-based estimators. In summary,  in the presence of omitted confounders, the novel estimation method produces consistent estimators and reliable CIs for the true EIN, NNE, and NNT. Additional simulation analysis for smaller causal effects and consequent higher values of EIN, NNE and NNT can be found in Appendix~\ref{app:sim2}.

\begin{table}[ht]
	\centering
	\begin{tabular}{ll|lll|lll}
		Model &  &  &  logit & &  &  probit &\\
		\hline
		$n$ & Measure &   EIN (4.18) & NNE (5.60) & NNT (4.65) &            EIN (2.81) & NNE (3.41) & NNT (3.02) \\ 
		\hline
		500& Coverage &  0.961 & 0.927 & 0.935 &       0.998   &  0.973   &  0.991 \\ 
		& MCSE   &     0.097  & 0.116 &  0.102  &                  0.025 &  0.029  & 0.026 \\ 
		& Av. bias  &        1.548  & 1.969  & 1.658        &              0.480  & 0.597  & 0.501 \\ \hline 
		1000& Coverage  &  0.945   &  0.932   &  0.933 &      0.998  &   0.959  &   0.984\\  
		& MCSE   &      0.046 & 0.060 &  0.050  &                0.015  & 0.018 &  0.015 \\ 
		& Av. bias  &       0.949  & 1.258  & 1.030       &                 0.346  & 0.424  & 0.358\\ \hline
		2000&  Coverage  & 0.957  &   0.944   &  0.949 &       0.992  &   0.962  &   0.987\\
		& MCSE   &        0.026  & 0.034  & 0.028   &                  0.010  & 0.012 & 0.010 \\ 
		& Av. bias  &    0.602  & 0.796  & 0.652      &           0.249  & 0.288  & 0.248\\  \hline
		4000&Coverage & 0.958    & 0.948  &    0.950 &       0.984  &   0.954  &   0.983 \\  
		& MCSE   &         0.016 &  0.021  & 0.017      &                0.006  & 0.008  &  0.007 \\ 
		& Av. bias  &    0.397 & 0.526 & 0.428     &           0.167  & 0.186   & 0.163 \\
	\end{tabular}
	\caption{\small{$95\%$-level CIs for the marginal EIN, NNE, and NNT, the estimators' Monte Carlo standard errors (MCSE), and average bias (Av. bias) as a function of the sample size~$n=500, 1000, 2000, 4000$. The number of iterations for each sample size is $m = 1000$. The strength of the IV was measured as the mean values of the Wald statistic of the IV regression coefficient in the exposure model, i.e., $E[Z_{stat.}]$ for $\gamma_1$ in~\eqref{dgp:Z&A}. The estimated strength of the instrument was $11.90$, $16.95$, $24.00$, and $33.94$ for each sample size, respectively. ``Bread'' matrices with condition number of $\ge 10 ^ {12}$ were excluded from further analysis since they produce numerically singular covariance matrices and do not allow for the construction of the analytical CIs.   }}\label{tb:coverage_rate}
\end{table}

\section{Real world data example}\label{sec6}
To illustrate our novel estimation method in a real-world scenario, we use the vitamin~D data available in the ivtools R-package. These publicly available data are a modified version of the original data from a cohort\cite{skaaby2013vitamin} study on vitamin~D status causal effect on mortality rates previously used by Sjölander \& Martinussen.\cite{sjolander2019instrumental} Vitamin~D deficiency has been linked with several lethal conditions such as diabetes, cancer, and cardiovascular diseases. However, vitamin~D status is also associated with several behavioral and environmental factors, such as season and smoking habits, that may result in biased estimators when using standard statistical analyses to estimate causal effects. 

Mendelian randomization\cite{davey2003mendelian, smith2004mendelian} is a method whose principles were  introduced originally  by Katan\cite{katan1986apoupoprotein} in a medical context. Subsequently, Youngmen et al.\cite{youngman2000plasma}  introduced this method in the context of epidemiological studies and also coined the aforementioned term. The underlying principle of the method is to use genotypes as IVs to estimate the causal effect of phenotype on disease-related outcomes. The population distribution of genetic variants is assumed to be independent of behavioral and environmental factors that usually confound the effect of exposure on the outcome. The process governing the distribution of genetic variants in the population resembles the randomization mechanism in RCTs. 

In our example, the phenotype is vitamin~D baseline status, the outcome is survival at the follow-up endpoint, and the genotype is mutations in the filaggrin gene. These mutations are associated with a higher serum vitamin~D concentration. The prevalence of this mutation is estimated to be $8\%-10\%$ in the northern European population. We used the modified version of data obtained in the Monica10 population-based study. This is a $10$-years follow-up study started in 1982-1984 that initially included an examination of $3,785$ individuals of Danish origin. In the follow-up study of 1993-1994, the participation rate was about~$70\%$. It resulted in data that contained a total of $2,656$ participants, where $2,571$ were also available in the modified data after the removal of cases that had missing information on filaggrin and or on vitamin~D status.\cite{martinussen2019instrumental, skaaby2013vitamin}  These data consisted of~$5$ variables: age (at baseline), filaggrin (a binary indicator of whether filaggrin mutations are present), vitd (vitamin D level at baseline as was assessed by serum 25-hydroxyvitamin~D 25-OH-D(nmol/L) concentration on serum lipids), time (follow-up time), and death (an indicator of whether the subject died during follow-up). 

This analysis considers only the necessary variables for the required estimators. We use the binary indicator of survival at the follow-up endpoint as the outcome~$I$ and the binary indicator of the presence of the filaggrin gene mutation as the IV~$Z$. For the binary exposure~$A$, the  vitamin~D status at baseline (vitd) was dichotomized at a threshold serum level of~$30$ ng/mL, defined as the bottom end of the acceptable range of vitamin~D for skeletal health.\cite{heaney2011iom} Values higher (or equal)~$30$ were defined as~$1$  to indicate an exposure, and~$0$ otherwise to indicate non-exposure. The estimated models were the double logit and the double probit models. In particular, the explicit form of the assumed GSMM are as presented in eq.~\eqref{eq:logit_causal_model} and eq.~\eqref{eq:probit_causal_model}, and the form of the assumed association models for the observed outcome are as presented in eq.~\eqref{eq:logit_assoc_mod}, and eq.~\eqref{eq:probit_assoc_mod}, for the logit and probit models, respectively. Namely, for each model, the core estimands are two causal parameters $\psi = (\psi_0, \psi_1)^T$, four association model parameters $\beta_I = (\beta_0, \beta_1, \beta_2, \beta_3)^T$, and one IV model parameter~$\pi_Z$. The indices of interest (EIN, NNE, and NNT) are functions of these parameters.   The populationwise NNT in such a case is the average number of randomly drawn individuals whose vitamin~D level needs to be increased to the normal range (above $30$ ng/mL) to prevent additional death during the follow-up period. Analogically, the NNE is the average number of randomly drawn individuals from the unexposed group whose vitamin~D level needs to be elevated to the normal range to prevent one additional death during the follow-up period. The EIN is defined similarly to NNE for the exposed group. The marginal probability~$\pi_Z$ was estimated using the sample mean of the filaggrin gene mutation indicator. We used both the double logit and double probit models to estimate the EIN, NNE, and NNT. 


The EIN point estimators with the corresponding $95\%$-level CI were $1.53$ $[1.16, 1.91]$, and $1.51$ $[1.12, 1.90]$, for the double logit and double probit models, respectively. There was no solution for~$\psi_0$ under these two models, therefore, we were unable to compute the corresponding NNE and NNT estimators. 
Namely, according to the results obtained from the novel estimation method, increasing the vitamin~D level to the normal range in the exposed group (i.e., the group with normal vitamin~D levels) is highly effective for preventing fatal outcomes.  
The absence of a solution for~$\psi_0$ can result from various reasons, where the most probable is insufficient data. Namely, the unexposed group constitutes only about~$7\%$ of the cohort, combined with a low filaggrin gene mutation prevalence in this group ($<3\%$), which may result in insufficient sample size for reliable estimation of $\psi_0$. A possible consequence of the insufficient sample size is the weak instrument (the estimated Wald statistic of the filaggrin gene mutation regression coefficient in the exposure model was $Z_{stat} = 2.446$). Another possible reason includes the misspecification of either the association (outcome) or the causal model for the unexposed group. Yet since most of the observations ($93\%$) are from the exposed group, assuming that the true NNE is reasonably small, we can deduce that the NNT's true value is close to the EIN value. Namely,  increasing the vitamin~D level to the normal range in the whole population is also expected to be highly effective for fatal outcomes prevention. The source code for this analysis is available in the author's GitHub repository.\footnote{Vitamin D data analysis source code: \url{https://github.com/vancak/nne_iv/tree/main/VitD_analysis}}

\section{Summary and Conclusions}\label{sec7}
In observational studies,  the exposed group characteristics may substantially differ from the unexposed. To address this issue, groupwise efficacy indices were defined: the EIN for the exposed group and the  NNE for the unexposed. In such studies, the group allocation is typically affected by confounders. In many studies, the measured confounders cannot be assumed sufficient for confounding control. Therefore, using the available estimation methods to estimate the EIN, NNE, or NNT will result in inconsistent estimators. 
Using Rubin's potential outcomes framework, this study explicitly defines the NNT and its derived indices, EIN and NNE, as causal measures. Then, we introduce a novel method that uses IVs to estimate the three aforementioned indices in observational studies where the omission of unmeasured confounders cannot be ruled out. We present two analytical examples – double logit and double probit models. Next, a corresponding simulation study is conducted. The simulation study illustrates the improved performance of the new estimation method compared to the available methods in terms of consistency and CIs empirical coverage rates. Finally, a real-world example based on a study of vitamin~D deficiency effects on mortality rates is presented. In this example, we evaluate the efficacy of increasing vitamin~D to normal levels to prevent fatal outcomes. The new estimator suggests that increasing vitamin~D level above 30 ng/mL among the exposed group is highly efficient in preventing fatal outcomes. 

The novel estimators' statistical consistency relies on the instrumental variable's validity. One may distinguish between the relevance assumption, i.e., the association between the IV and the exposure, and the exogeneity with the exclusion assumptions that are captured by~$I_a \indep Z|L, a \in \{0,1\}$.  The distinction is made since the relevance assumption is usually not completely violated as, in such a case, the variable cannot be used as an instrument in the first place. Frequently, the practical problem, such as the one we encountered in the Vitamin~D example, is an insufficient strength of the instrument, i.e., a weak instrument. Weak instrumental variables are instruments with low explanatory power for the model's endogenous variable (i.e., the exposure indicator). Staiger and Stock\cite{staiger1994instrumental} suggested a practical rule of thumb for linear instrument-exposure models: if the F-statistic for the first-stage instrument-exposure model does not exceed~$10$, the instrument is considered weak. In non-linear models, especially in saturated models with binary instruments and binary exposure, one can assess the instrument's strength by computing the probability of compliance. In general cases, one can use measures based on the magnitude of the Wald test\citep{menard2004six} for the instrument regression coefficient or the magnitude of pseudo-R-squared statistics.\citep{cameron1997r} Using weak instruments can pose challenges in estimating causal effects by exacerbating biases even in large samples, increasing the estimators’ sample variance and thus resulting in very wide and unreliable CIs.\cite{staiger1994instrumental, burgess2017review} Remedies for obtaining reliable confidence intervals and hypothesis testing with weak instruments include using robust methods based on the Anderson-Rubin or conditional likelihood ratio statistics.\citep{anderson1949estimation,moreira2003conditional} 
The problem of weak instruments differs conceptually from the exogeneity with the exclusion assumption violations. In the latter case, the induced bias does not vanish asymptotically and increasing the sample size cannot reduce it; thus, unlike for the weak instruments, robust methods cannot be constructed without additional assumptions and further modelling. In such scenarios, the resulting estimators are statistically inconsistent. Possible remedies include finding different candidate instruments or modelling the violation structure. However, parametric models of the violation structure do not guarantee the identification of these parameters;\cite{vancak2023sensitivity} hence, sensitivity analysis has to be performed. Nevertheless, even a sensitivity analysis is not a simple task since specifying a plausible interval for the sensitivity parameters may be challenging, too, as it relies on subject-matter knowledge and a solid understanding of causal inference methodology. All these considerations are possible directions for prospective research on robust inference methods and sensitivity analysis.

Additionally, in the examined models, we considered only binary observed outcomes. However, in many studies, the observed outcome is non-binary, and the dichotomization is performed using a certain threshold, as we did in the vitamin D example. This practice has disadvantages - dichotomization results in loss of Fisher information and, therefore, a decrease in statistical power.\cite{senn2003disappointing, fedorov2009consequences}  Although the illustrated theoretical setup is non-parametric, the examined causal and association models are semi-parametric. These types of models are attractive since they simplify the estimation procedure. However, such models are sensitive to model misspecification as we need to specify three different models: (1) one for the instrumental variable, (2) one for the association (outcome) model, and (3) one for the GSMM. This results in three sets of parameters that need to be estimated and three different models that can be misspecified. Therefore, non-parametric estimation might also serve as a possible direction for future research. 

The main  contribution of this study is by providing explicit causal formulation of the EIN, NNE, and NNT indices and a comprehensive theoretical framework for their point and interval estimation using the G-estimators in observational studies with unmeasured confounders. Future research direction may focus either on applications or extensions of the novel method to new domains.   



\appendix

\section{APPENDIX}

\subsection{Proof of unbiasedness of the G-estimators estimating equations}\label{app:cons_g} 
Without the loss of generality, we present the proof for the exposed group, $a=1$. The proof for the unexposed group can be readily obtained using analogical steps for~$a=0$. Assume that~$\xi$ is a link function for the structural mean model~\eqref{eq:gsmm} and let~$\eta$  be the link function for the association model~$\eta([E|Z, L, A;\beta_I])$. Let the estimating equation for the G-estimator of~$\psi_1$ be as defined in eq.~\eqref{eq:dh}. Therefore,
\begin{align}
	E[ D_1(Z, L; \pi_Z) h(1; Z, L, 1,  \beta_I, \psi_1) ]  &
	= E[ D_1(Z, L; \pi_Z) \xi^{-1} \left( \eta(E[I|Z,L,A=1; \beta_I]) - m_1^T(L)  \psi_1   \right) ] \label{app_gsmm} \\ 
	& = E[ D_1(Z, L; \pi_Z)\xi^{-1} \left( \eta(E[I_1|Z,L,A=1; \beta_I]) - m_1^T(L)  \psi_1   \right)   ] \label{app_cons} \\
	& = E[ D_1(Z, L; \pi_Z) E[I_0|Z, L, A=1]  ]  \label{counter_predict}\\
	& = E[ E[ D_1(Z, L; \pi_Z) I_0|Z, L, A = 1  ]] \label{app1} \\
	& = E[ D_1(Z, L; \pi_Z) I_0 ]\label{app2} \\ 
	& = 0. \nonumber
\end{align}
Eq.~\eqref{app_gsmm} is a direct consequence of the assumed GSMM structure in eq.~\eqref{eq:gsmm}. Eq.~\eqref{app_cons} stems from the consistency assumption.\cite{didelez2010assumptions} Eq.~\eqref{counter_predict} is obtained by plugging-in eq.~\eqref{eq:counter_predict} for $a=1$. Eq.~\eqref{app1} holds since the potential outcome~$I_0$ is independent of the observed allocation~$A$. The last step in eq.~\eqref{app2} stems from the validity of the IV. Namely, the estimating equations are unbiased as long as the IV~$Z$ satisfies conditions of a valid IV, i.e., $I_0 \indep Z | L$.


\subsection{The double probit model example}\label{app:double_probit_example}
Assume a binary outcome $I \in \{0, 1\}$, a binary exposure $A \in \{0, 1\}$, and a binary instrument~$Z \in \{0, 1\}$. Assume there are no measured confounders, i.e., $L = \emptyset$ and $m(L) = 1$.  For the probit link function~$\xi$, we define the GSMM~\eqref{eq:gsmm} as
\begin{align}\label{eq:probit_causal_model}
	\Phi ^{-1} \left(E[ I_1  = 1 | Z, A = a ] \right) - \Phi^{-1} \left( E[ I_0  = 1 | Z, A = a ] \right) = \psi_a, \quad a \in \{0, 1\},
\end{align}
where $\Phi^{-1}(x)$ is the inverse of the standard normal random variable's cumulative distribution function~$\Phi(x) = \int_{-\infty}^{x} (2\pi)^{-1/2} \exp\{ - s ^ 2 / 2\}ds$. Additionally, assume the following saturated probit model for the observed outcome
\begin{align}\label{eq:probit_assoc_mod}
	\Phi^{-1} \left(  E[I|Z, A ; \beta_I] \right) = \beta_0 + \beta_1A + \beta_2 Z + \beta_3 ZA.	
\end{align} 
Using the GSMM and the probit association model, we can express explicitly the general conditional exposure benefit~\eqref{eq:exp_prob} as a function of~$\beta_I$ and~$\psi$
\begin{align}\label{eq:probit_exp_prob}
	p_b(Z,  A, \beta_I, \psi ) 
	&= 
	\Phi 
	\left(
	\beta_0 + \beta_1A + \beta_2Z + \beta_3ZA	
	+  \psi_{0}(1-A) )
	\right)\\
	& -
	\Phi 
	\left(
	\beta_0 + \beta_1A + \beta_2Z + \beta_3ZA		
	- \psi_{1} A 
	\right). 	\nonumber 
\end{align}
To compute the populationwise NNT, we apply the function~$g$ to the marginal exposure benefit~$p_b = E[p_b(Z, A,  \beta_I,\psi)]$~\eqref{eq:marginal_pb}, i.e.,  $\text{NNT} = g(p_b)$. If we set~$A$ to~$0$ in eq.~\eqref{eq:probit_exp_prob} we obtain the conditional exposure benefit~\eqref{eq:cond_p_b(a)} for the unexposed $a=0$ as a function of $\beta_I$ and $\psi_0$ 
\begin{align}\label{eq:probit_p_unexposed}
	p_b(0; Z, \beta_I, \psi_0) =  \Phi ( \beta_0 + \beta_2Z + \psi_0)
	- 
	\Phi( \beta_0 + \beta_2Z),
\end{align} 
To compute the NNE, we apply the function~$g$ to the marginal exposure benefit for the unexposed $p_b(0) = E[p_b(0; Z, \beta_I, \psi_0)|A=0]$~\eqref{eq:marginal_pb(a)}, i.e., $\text{NNE} = g(p_b(0))$.   Analogically, for the conditional exposure benefit~\eqref{eq:cond_p_b(a)} for the exposed $a=1$, we set~$A$ to~$1$ in eq.~\eqref{eq:logit_exp_prob}
\begin{align}\label{eq:probit_p_exposed}
	p_b(1; Z, \beta_I, \psi_1)  & = \Phi(\beta_0 + \beta_1 + \beta_2 Z + \beta_3 Z)
	- \Phi(\beta_0 + \beta_1 + \beta_2Z + \beta_3 Z - \psi_1).
\end{align}
To compute the EIN, we apply the function~$g$ to the marginal exposure benefit for the exposed~$p_b(1) = E[p_b(1; Z, \beta_I, \psi_1)|A=1]$~\eqref{eq:marginal_pb(a)}, i.e., $\text{EIN} = g(p_b(1))$.

\subsection{Explicit form of valid IV conditions}\label{app:valid_iv}
\subsubsection{The double logit model}
Let the GSMM be defined as in eq.~\eqref{eq:logit_causal_model}, and the marginal distribution of the IV~$Z$ and the conditional distribution of the exposure~$A$ are as specified in~\eqref{dgp:Z&A}. For a double logit model, the explicit form of eq.~\eqref{eq:valid_iv_binaryZ} for the potential outcome~$I_0$ is 
\begin{align}\label{eq:logit_IV_Iu}  
	&\text{expit}(\beta_0 + \beta_2 ) ( 1 -  \text{expit}(\gamma_0 + \gamma_1) )
	+
	\text{expit}(\beta_0 + \beta_1 + \beta_2 + \beta_3 - \psi_1) \text{expit}(\gamma_0 + \gamma_1) \nonumber\\
	&=
	\text{expit}(\beta_0) ( 1 -  \text{expit}(\gamma_0) )
	+
	\text{expit}(\beta_0 + \beta_1 - \psi_1) \text{expit}(\gamma_0), 
\end{align}
and for the potential outcome~$I_1$ is
\begin{align}\label{eq:logit_IV_Ie}
	&  
	\text{expit}(\beta_0 +  \beta_2 + \psi_0) (1 - \text{expit}(\gamma_0 + \gamma_1) )
	+
	\text{expit}(\beta_0 + \beta_1 +  \beta_2 + \beta_3 ) \text{expit}(\gamma_0 + \gamma_1) \nonumber\\
	&= 
	\text{expit}(\beta_0 + \psi_0) ( 1 -  \text{expit}(\gamma_0) ) 
	+ 
	\text{expit}(\beta_0 + \beta_1) \text{expit}(\gamma_0) .
\end{align}
\subsubsection{The double probit model}
Let the GSMM be defined as in eq.~\eqref{eq:probit_causal_model}, and the function~$\xi$ be the probit~$\Phi^{-1}$ function. The explicit forms of the exposure benefits are given in eq.~\eqref{eq:probit_p_unexposed} and \eqref{eq:probit_p_exposed}.  The explicit forms of eq.~\eqref{eq:valid_iv_binaryZ} for the potential outcomes~$I_0$ and $I_1$ are obtained by replacing the expit function with the inverse probit function~$\Phi$ in equations~\eqref{eq:logit_IV_Iu} and~\eqref{eq:logit_IV_Ie}, respectively.

\subsection{Simulation study, Step~2: Estimation}\label{app:sim_step2}
This subsection presents the explicit form of the vector-valued function $\mathbf{Q} ( Z, A; \theta )$ components. The vector of unbiased estimating functions~\eqref{eq:est_fun_b} of~$\beta_I$ and $\pi_Z$ consists of the score functions of the association model~\eqref{dgp:outcome_model} and the binary instrument model~\eqref{dgp:Z&A}, namely, 
\begin{align*}
	\mathbf{S}(Z, A; \beta_I, \pi_Z) = 
	\begin{pmatrix}
		(I - \xi^{-1}(\beta_0 + \beta_1 A + \beta_2 Z + \beta_3 AZ ))\\
		(I - \xi^{-1}(\beta_0 + \beta_1 A + \beta_2 Z + \beta_3 AZ ))A\\
		(I - \xi^{-1}(\beta_0 + \beta_1 A + \beta_2 Z + \beta_3 AZ ))Z\\
		(I - \xi^{-1}(\beta_0 + \beta_1 A + \beta_2 Z + \beta_3 AZ ))AZ\\
		(Z-\pi_Z)
	\end{pmatrix}.
\end{align*}
The vector-valued function $\mathbf{Dh}(Z, A; \beta_I, \pi_Z, \psi)$ as in eq.~\eqref{eq:dh} of the two estimating functions for the causal parameters~$\psi$ is
\begin{align*}
	\mathbf{Dh}(Z, A; \beta_I, \pi_Z, \psi ) = 
	\begin{pmatrix}
		(Z-\pi_Z) \xi^{-1}\left(\beta_0 + \beta_1 A + \beta_2 Z + \beta_3 AZ + \psi_0(1-A) \right)\\
		(Z-\pi_Z) \xi^{-1}\left(\beta_0 + \beta_1 A + \beta_2 Z + \beta_3 AZ - \psi_1A \right)\
	\end{pmatrix}.
\end{align*}
The vector-valued function $ \mathbf{p}\left(Z, L, A; \beta_I, \psi, p_b(0), p_b(1), p_b \right)$ as in eq.~\eqref{eq:est_fun_p} consists of the three estimating equations for the exposure benefits
\begin{align*}
	& \mathbf{p}\left(Z, L, A; \beta_I, \psi, p_b(0), p_b(1), p_b \right)\\ 
	&= 
	\begin{pmatrix}
		\left( \xi^{-1}\left(\beta_0 + \beta_2 Z  + \psi_0 \right) - \xi^{-1}\left(\beta_0 +  \beta_2 Z  \right) - p(0)\right)(1-A)\\
		\left( \xi^{-1}\left(\beta_0 + \beta_1  + \beta_2Z  + \beta_3 Z \right) - \xi^{-1}\left(\beta_0 + \beta_1  + \beta_2Z  + \beta_3 Z - \psi_1   \right) - p(1)\right)A\\
		\xi^{-1}\left(\beta_0 + \beta_1 A + \beta_2 Z + \beta_3 AZ + \psi_0(1-A) \right) - \xi^{-1}\left(\beta_0 + \beta_1 A + \beta_2 Z + \beta_3 AZ - \psi_1 A \right) - p_b\\
	\end{pmatrix}.
\end{align*}
In all components of the estimating function $\mathbf{Q}(Z, A; \theta)$, $\xi^{-1}$ is the expit and the inverse probit $\Phi$ functions for the double logit and double probit models, respectively. Notably, the vector valued function~$ \mathbf{g}\left(p_b(0), p_b(1), p_b, \text{NNE, EIN, NNT}\right)$ is independent of the data, and thus its form remains the same as in its definition in~\eqref{eq:est_fun_g}.

\subsection{Simulation study -  Setting II}\label{app:sim2}
This simulation setting is analogical to the simulations in Section~\ref{sec5}, except for smaller values of the causal parameters $\psi$ and higher consequent values of the corresponding EIN, NNE and NNT measures. Namely, we use the double logit and double probit models described in the simulation setup in Subsection~\ref{subsec:sim_setup} with the same sample sizes. This setting can serve as a preliminary sensitivity analysis of the methodology for combining small sample sizes with small causal effects.
\begin{figure}
	\begin{subfigure}{0.5\textwidth}
		\centering
		\includegraphics[width=1\linewidth]{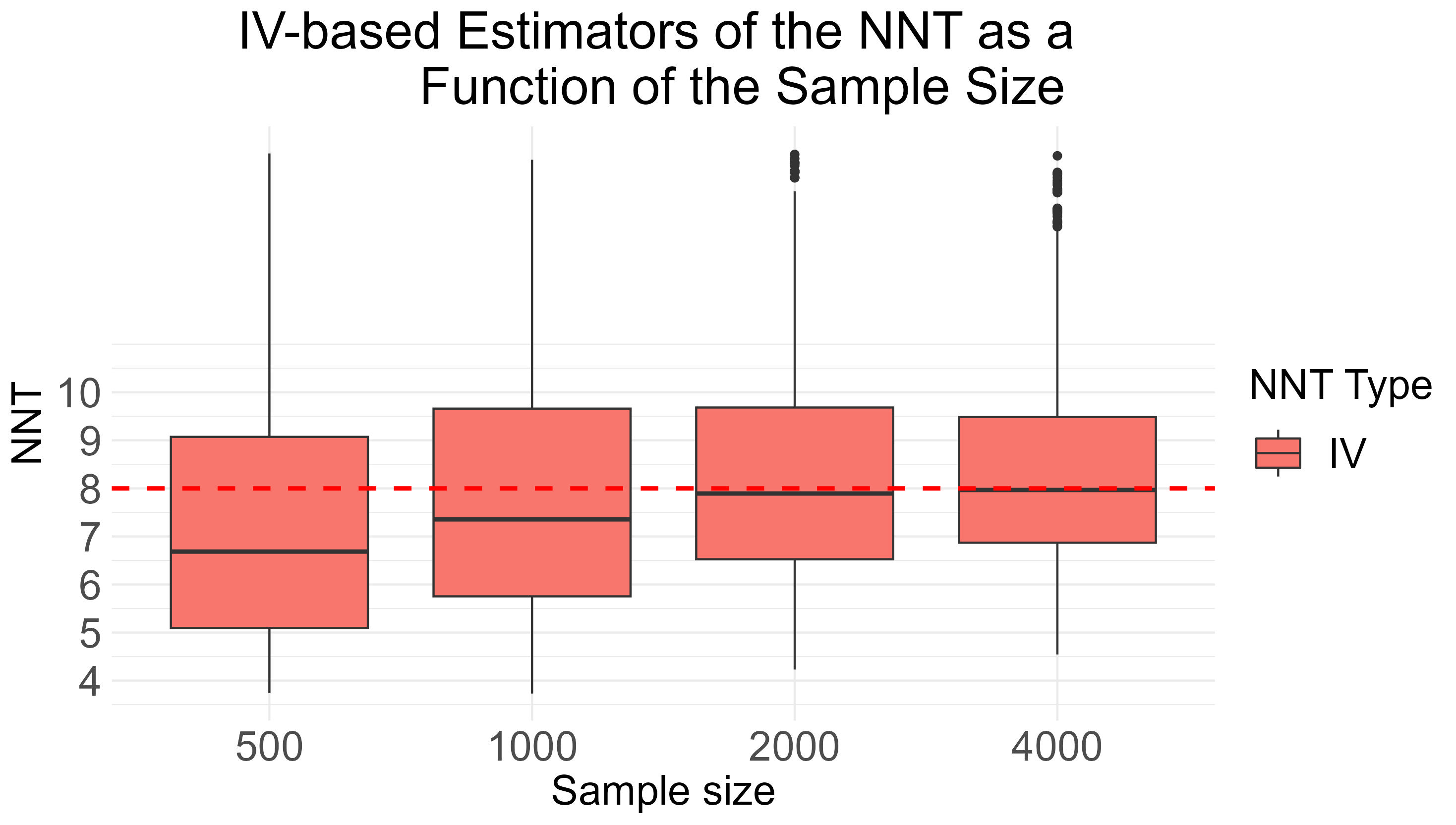}
		\caption{Double logit model. True NNT$= 8.00$.}
	\end{subfigure}%
	\begin{subfigure}{0.5\textwidth}
		\centering
		\includegraphics[width=1\linewidth]{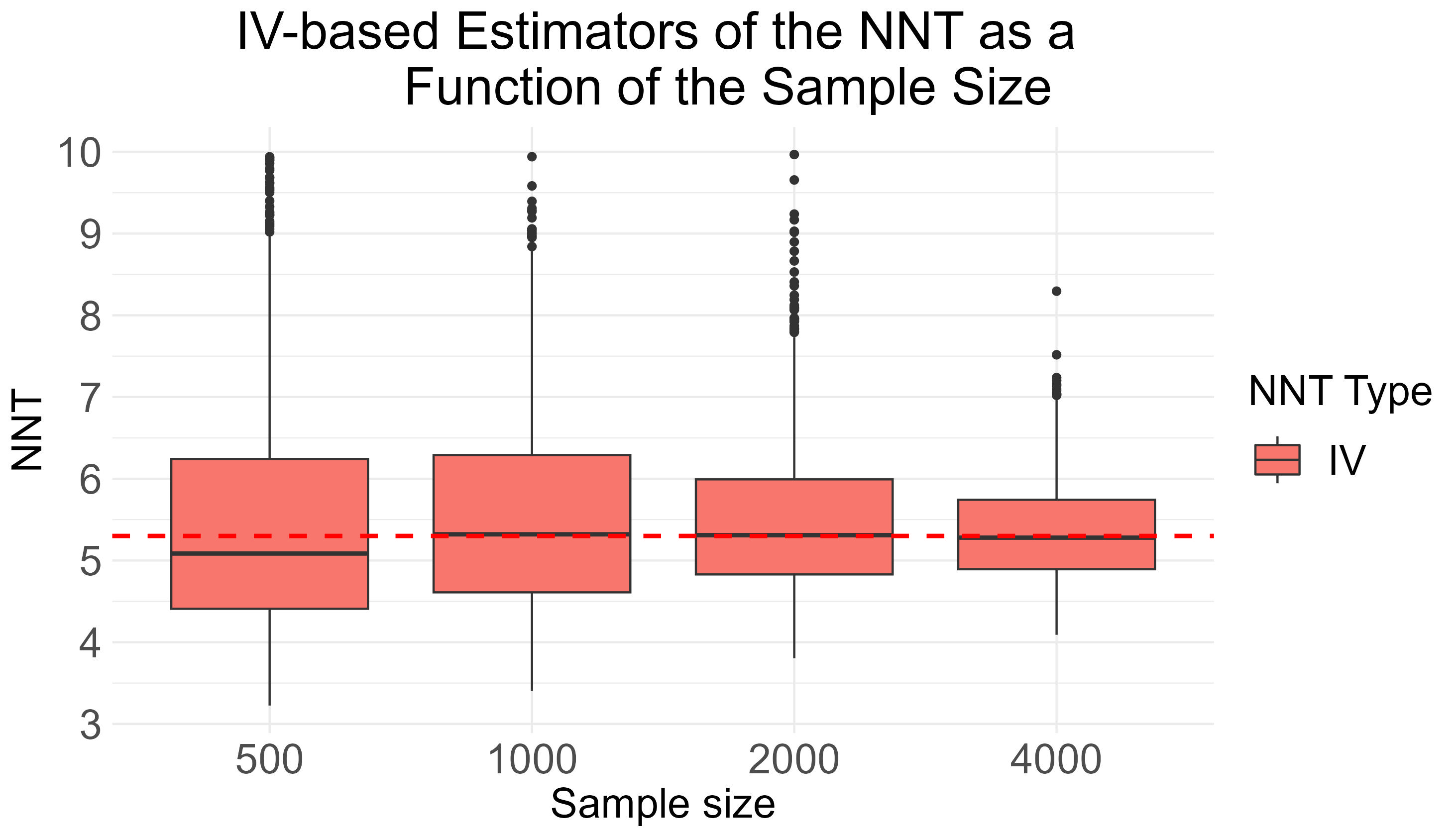}
		\caption{Double probit model. True NNT$= 5.30$.}
	\end{subfigure}
	\caption{Logit and probit models examples: Binary outcome with logit and probit causal models, respectively. The true causal parameters for both models are $\psi = (0.5, 1)^T$. The true populationwise NNTs are $8.00$, and $5.30$, respectively.  The red boxplots denote the IV-based estimators of the NNT, which are based on G-estimators of~$\psi$.  The unadjusted estimators of the NNT were infinitely large (the estimated benefits were negative before the application of the function~$g$~\eqref{def.g}); therefore, they were omitted from the graph. The red dashed line denotes the true NNT value for each model. The calculations were repeated~$m=1000$ times for four different sample sizes: $n = 500, 1000, 2000, 4000$. For both models, the marginal $P(A=1)=0.6$, $\pi_Z = 0.5$, $\gamma = (-0.83, 3)^T$ and the marginal probability of the outcome is $P(I=1) = 0.3$.}\label{sim2:NNT}
\end{figure}

\begin{figure}
	\begin{subfigure}{0.5\textwidth}
		\centering
		\includegraphics[width=1\linewidth]{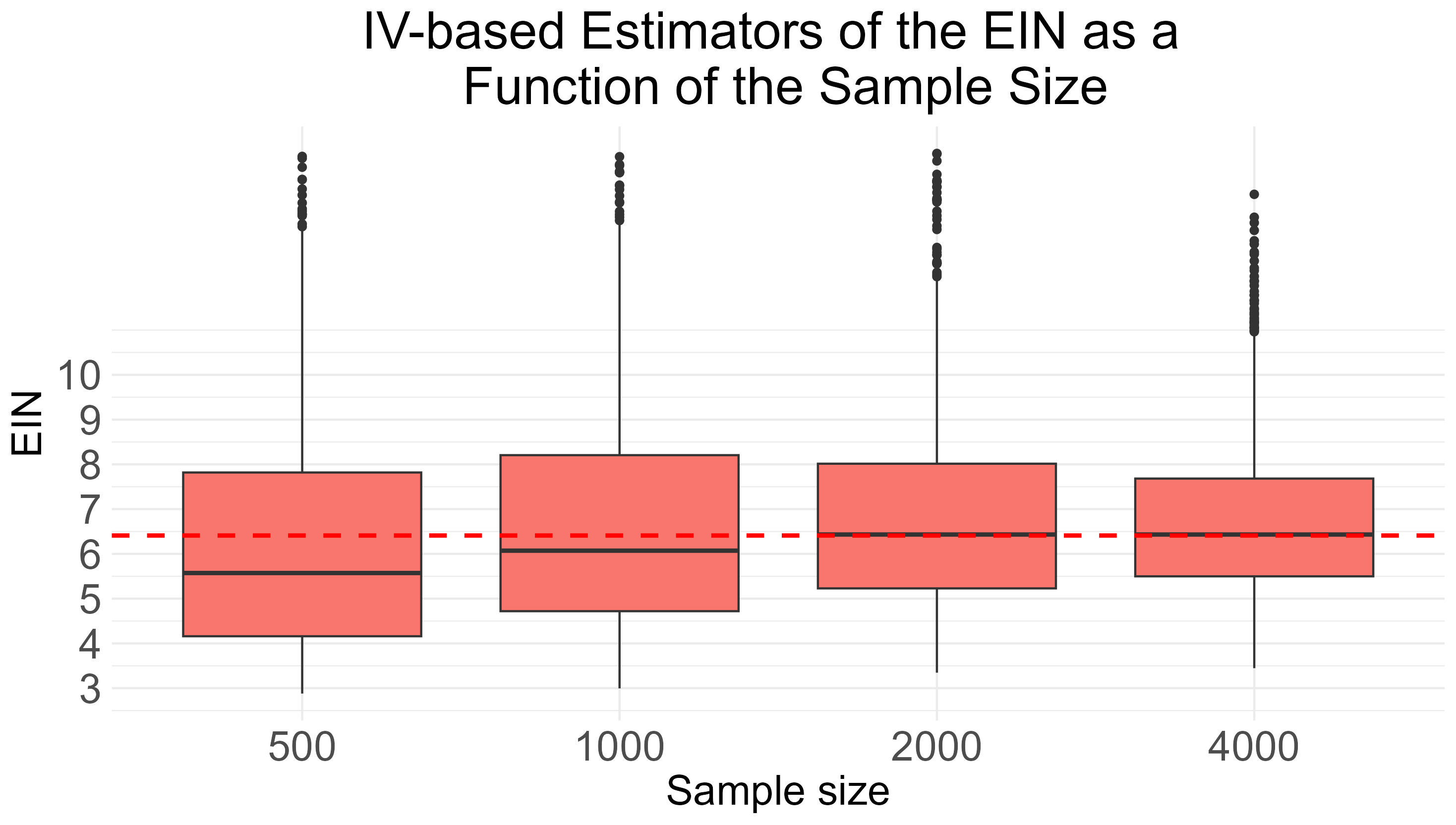}
		\caption{Double logit model. True EIN$= 6.41$.}
	\end{subfigure}%
	\begin{subfigure}{0.5\textwidth}
		\centering
		\includegraphics[width=1\linewidth]{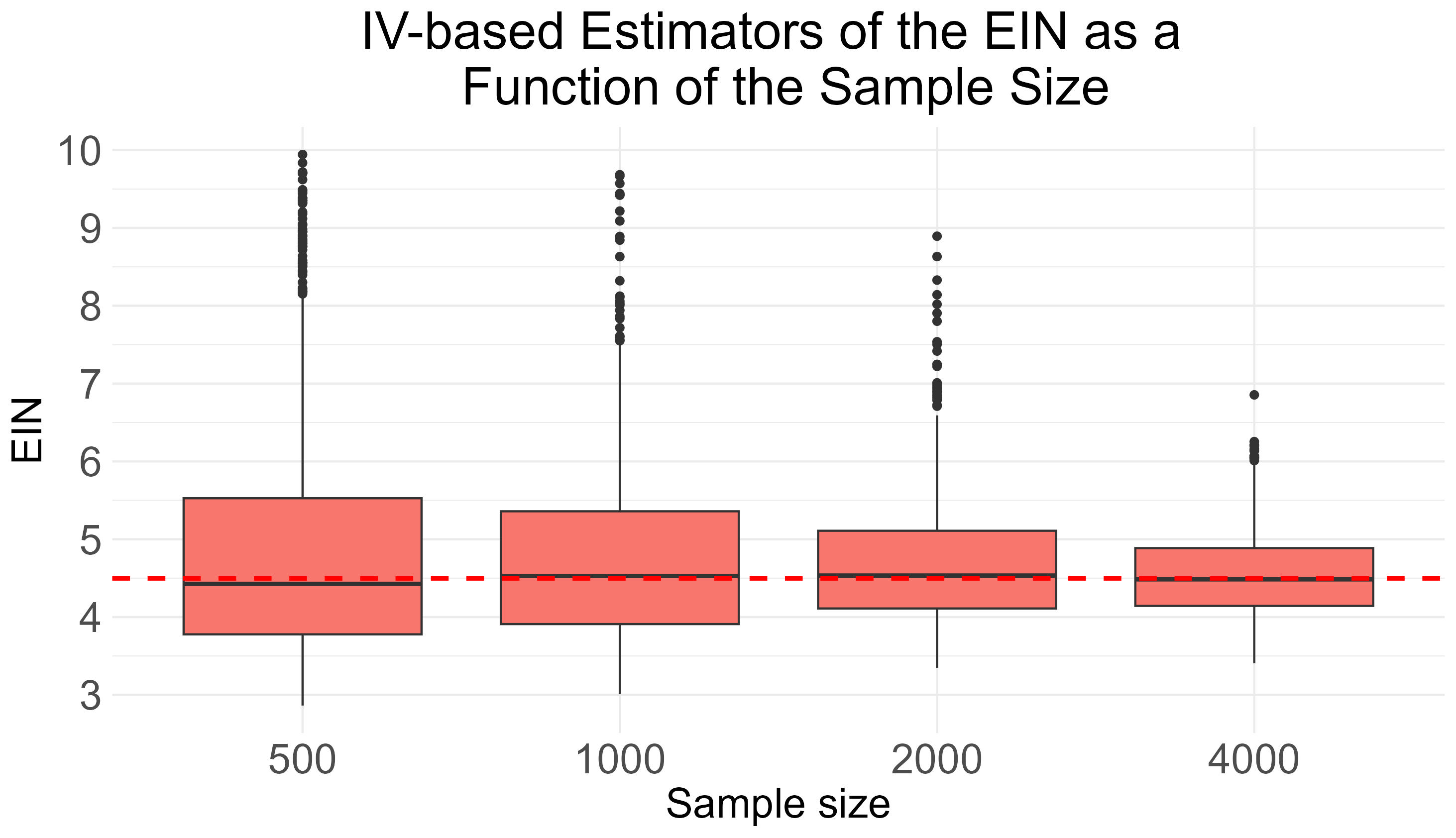}
		\caption{Double probit model. True EIN$= 4.50$.}
	\end{subfigure}
	\caption{\small{Double logit and probit models examples: Binary outcome with logit and probit causal models, respectively. The causal parameters are $\psi = (0.5, 1)^T$. The true EINs are~$6.41$ and~$4.50$, respectively.  The red boxplots denote the IV-based estimators of the EIN, which are based on G-estimators of~$\psi$. 
			The unadjusted estimators of the EIN were infinitely large (the estimated benefits were negative before the application of the function~$g$~\eqref{def.g}); therefore, they were omitted from the graph.  The red dashed line denotes the true EIN value in each model. The calculations were repeated~$m=1000$ times for four different sample sizes: $n = 500, 1000, 2000, 4000$. For both models, the marginal $P(A=1)=0.6$, $\pi_Z = 0.5$, $\gamma = (-0.83, 3)^T$ and the marginal probability of the outcome is $P(I=1) = 0.3$.}  }\label{sim2:EIN}
\end{figure}

\begin{figure}
	\begin{subfigure}{0.5\textwidth}
		\centering
		\includegraphics[width=1\linewidth]{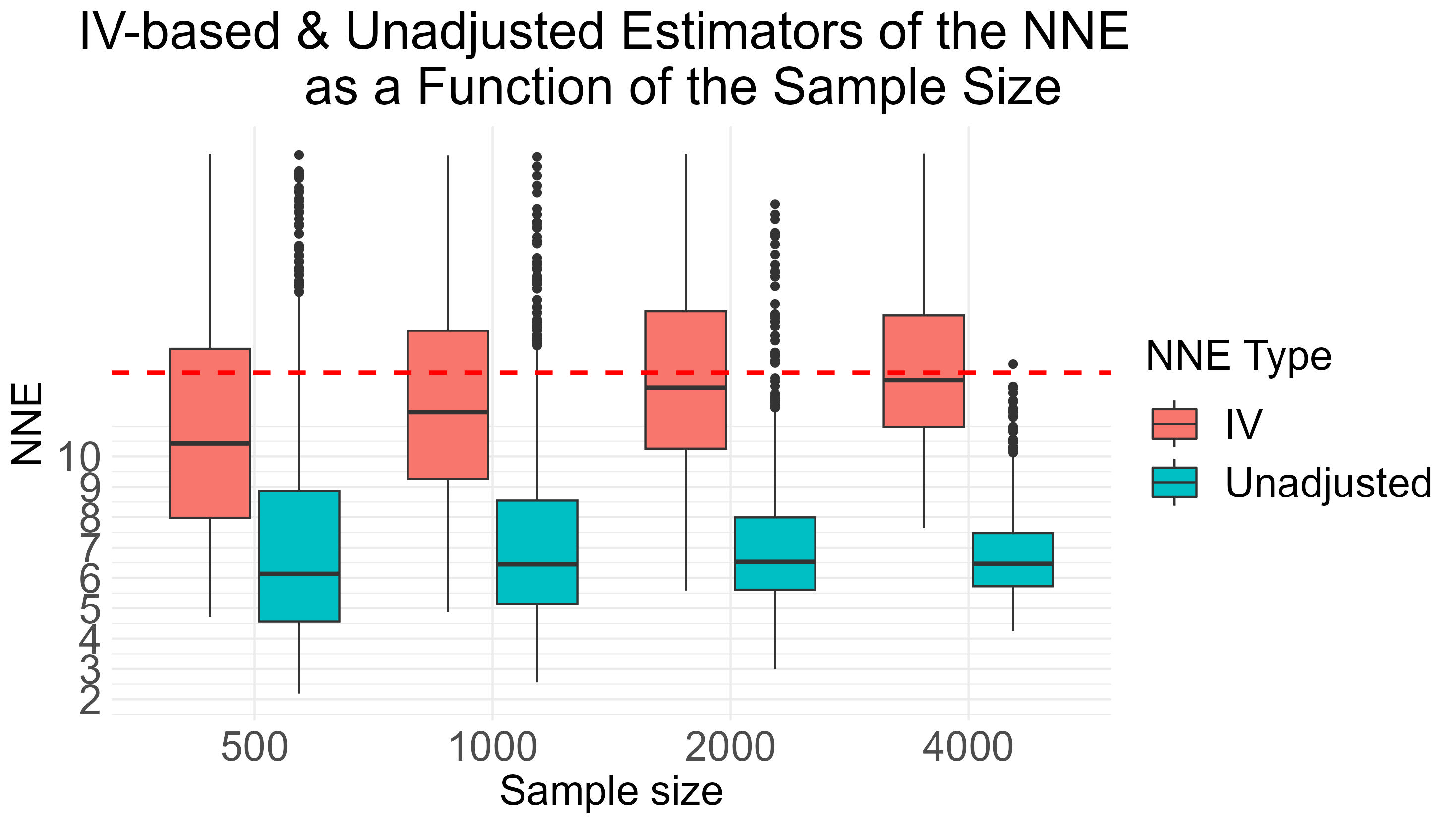}
		\caption{Double logit model. True NNE$= 12.77$.}
	\end{subfigure}%
	\begin{subfigure}{0.5\textwidth}
		\centering
		\includegraphics[width=1\linewidth]{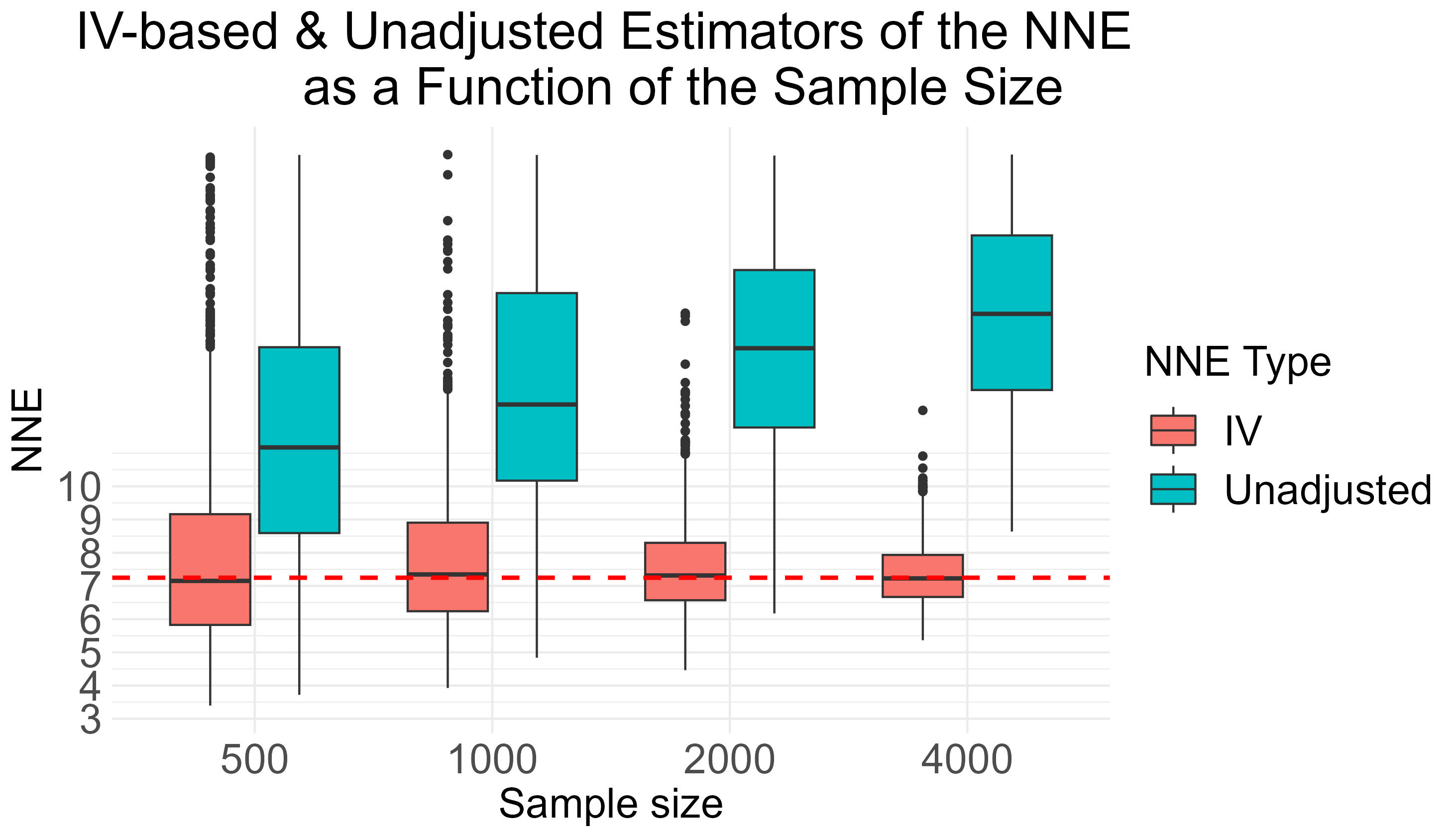}
		\caption{Double probit model. True NNE$= 7.25$.}
	\end{subfigure}
	\caption{
		\small{Double logit and probit models examples: Binary outcome with logit and probit causal models, respectively. The true causal parameters for both models are $\psi = (0.5, 1)^T$. The true populationwise NNEs are $12.77$, and $7.25$, respectively.  The red boxplots denote the IV-based estimators of the NNE, which are based on G-estimators of~$\psi$. The blue boxplots denote the unadjusted estimators of the NNE. The red dashed line denotes the true NNE value for each model. The calculations were repeated~$m=1000$ times for four different sample sizes: $n =  500, 1000, 2000, 4000$. For both models, the marginal $P(A=1)=0.6$, $\pi_Z = 0.5$, $\gamma = (-0.83, 3)^T$ and the marginal probability of the outcome is $P(I=1) = 0.3$.}  }\label{sim2:NNE}
\end{figure}

\begin{table}[ht]
	\centering
	\begin{tabular}{ll|lll|lll}
		Model &  &  &  logit & &  &  probit &\\
		\hline
		$n$ & Measure &   EIN (6.41) & NNE (12.77) & NNT (8.00) &            EIN (4.50) & NNE (7.25) & NNT (5.30) \\ 
		\hline
		500& Coverage &  0.957    & 0.937 & 0.941 &       0.994   & 0.948 & 0.979 \\
		& MCSE   &     1.158  & 1.714  & 1.397  &                  0.125  & 0.210  & 0.148 \\ 
		& Av. bias  &        8.733       &  14.610         &   10.537        &              1.718       &      3.010     & 2.017 \\
		& $\%$ Inf. CIs  &     2.9$\%$          &  4.6$\%$         &  4.8$\%$    &       13.5$\%$    &      0.01$\%$  &      13.4$\%$  \\ \hline 
		1000& Coverage  &  0.953    & 0.946 & 0.948 &       0.989   & 0.955 & 0.977\\
		& MCSE   &       0.798  &1.371  &0.946  &               0.086 &  0.141  & 0.100 \\ 
		& Av. bias  &       6.052        &  11.099         &   7.306        &                 1.041    &        1.838   & 1.218\\
		& $\%$ Inf. CIs  &     0.9$\%$          &  1.1$\%$         &   1.2$\%$   &              8.1$\%$    &      0.01$\%$   &      7.9$\%$  \\ \hline
		2000&  Coverage  &  0.954    & 0.938 & 0.946 &       0.989   & 0.959 & 0.979\\
		& MCSE   &        0.466  & 0.879  & 0.571   &                 0.026  & 0.046  & 0.031 \\ 
		& Av. bias  &     3.173          &     5.970      &     3.867      &            0.619         &      1.076     & 0.719\\
		& $\%$ Inf. CIs  &     0.5$\%$          &  0.5$\%$         &   0.6$\%$        &          2.6$\%$           &  0$\%$       & 2.5$\%$ \\  \hline 
		4000&Coverage &  0.955   & 0.937 & 0.952 &       0.978   & 0.954 & 0.969 \\
		& MCSE   &         0.063  & 0.121  & 0.077      &                0.018 & 0.030  & 0.020 \\ 
		& Av. bias  &     1.381          &      2.604     &     1.679      &           0.438          &        0.742   & 0.507 \\
		& $\%$ Inf. CIs  &       0$\%$          &      0$\%$       &          0$\%$   &             0.05$\%$        &  0$\%$       & 0.04$\%$  \\
	\end{tabular}
	\caption{\small{Double logit and probit models examples: Empirical coverage rates (Coverage) of the sandwich-matrix-based $95\%$-level CIs for the marginal EIN, NNE, and NNT, the estimators' Monte Carlo standard errors (MCSE), average bias (Av. bias), and the percentage of non-informative extremely large (upper limit > $1000$) CIs ($\%$ Inf. CIs), as a function of sample size~$n=500, 1000, 2000, 4000$. The number of iterations for each sample size is $m = 1000$. The strength of the IV was measured as the mean values of the Wald statistic of the IV regression coefficient in the exposure model, i.e., $E[Z_{stat.}]$ for $\gamma_1$ in~\eqref{dgp:Z&A}. The estimated strength of the instrument was $11.90$, $16.95$, $24.00$, and $33.94$ for each sample size, respectively. ``Bread'' matrices with condition number of $\ge 10 ^ {12}$ were excluded from further analysis since they produce numerically singular covariance matrices and do not allow for the construction of the analytical CIs.   }}\label{tab:2}
\end{table}

In summary, a combination of small causal effects that lead to high EIN, NNE and NNT values may result in unstable estimators with inflated average bias. Graphical summary of the estimators' behaviour as a function of the sample size~$n$ can be found in Figures~\ref{sim2:EIN}, ~\ref{sim2:NNE}, and~\ref{sim2:NNT}. Table~\ref{tab:2} presents the empirical coverage rates of the $95\%$-level CI, the MCSEs and the average bias for the marginal EIN, NNE, and NNT in double logit and probit models as a function of the sample size~$n$. The proportion of non-informative extremely large CIs is also non-neglectable for the EIN and NNT in the double-probit model for the sample size of $500$.  These caveats are mitigated only for moderate-size sample sizes of $2000$ and higher. A possible direction for future research is comprehensive sensitivity analysis and the development of robust estimation methods for such scenarios.

\bibliographystyle{plain}
\bibliography{NNT_CAUSAL_BIB2}

\end{document}